\documentclass[12pt,preprint]{aastex}

\usepackage{lscape}
\usepackage{graphicx}
\usepackage{subfigure}
\usepackage{multirow}
\usepackage{amsmath}

\begin{document}
\def\etal{{et al.}}
\def\ibvs{{IBVS}}
\title{The first photometric investigation and spectroscopic analysis of two contact binaries: ASAS J124343+1531.7 and LINEAR 2323566}
\author{Xia, Qiqi\altaffilmark{1}, Michel, Raul\altaffilmark{2}, Li, Kai\altaffilmark{1}, Higuera, Jesus\altaffilmark{3}}
\altaffiltext{1}{Shandong Key Laboratory of Optical Astronomy and Solar-Terrestrial Environment, School of Space Science and Physics, Institute of Space Sciences, Shandong University, Weihai, Shandong, 264209, China (e-mail: kaili@sdu.edu.cn (Li, K.))}
\altaffiltext{2}{Observatorio Astron\'{o}mico Nacional, Instituto de Astronom\'{i}a, Universidad Nacional Aut\'{o}noma de M\'{e}xico, Apartado Postal 877, Ensenada, B.C. 22830, M\'{e}xico}
\altaffiltext{3}{NSF's National Optical-Infrared Astronomy Research Laboratory, 950 N. Cherry Ave., Tucson, AZ 85719, USA}

\begin{abstract}
The multi-color passband CCD light curves of ASAS J124343+1531.7 and LINEAR 2323566 were first obtained by the 0.84-m Ritchey-Chr\'{e}tien telescope with follow up observations by the WIYN 0.90m Cassegrain telescope. The data from the $Transiting\quad Exoplanet\quad Survey\quad Satellite\quad (TESS)$ of ASAS J124343+1531.7 was also applied for subsequent analysis. By analyzing the data through the W-D program, their mass ratios and fill-out factors were determined as 3.758, 1.438 and 31.8$\%$, 14.9$\%$, respectively. ASAS J124343+1531.7 is a W-subtype median contact binary, while LINEAR 2323566 is a W-subtype shallow contact binary, and the asymmetric light curves prove that they both have the O'Connell effect, which is generally explained by magnetic activity. The equivalent widths (EWs) of H$_\alpha$ lines were calculated, which show they certainly have magnetic activity. Moreover, LINEAR 2323566 has a stronger magnetic activity. The analysis of orbital period changes shows that ASAS J124343+1531.7 has a trend of secular period increase, which is generally explained by the mass transfer from the less massive to the more massive star. According to the estimated absolute parameters, their evolutionary states are discussed. The two components of ASAS J124343+1531.7 are both main sequence stars. While for LINEAR 2323566, the more massive star is a main sequence star, the less massive star has evolved out of main sequence and is over-luminous and over-sized.
\end{abstract}


\section{Introduction}
W UMa type stars are close binary systems usually consisting of two late-type stars. The two stars fill or over-fill their Roche lobes, and there exists a common convective envelope(CCE)~\citep{Lucy68a,Lucy68b} around two components. Due to the existence of a CCE, there may be mass transfer and energy exchange between two components~\citep[eg.,][]{Li15,Liao17,Lee18,Li18a}. The mass and angular momentum may also be lost from the binary system~\citep[eg.,][]{Van79,Rahu81,van82,Qian01a,Qian01b,Qian03}. Thus a series of physical processes will occur, such as the O'Connell effect~\citep{Oconnell51}, thermal relaxation oscillation (TRO)~\citep[eg.,][]{Lucy76,Flannery76,Robertson77}, orbital period change, short period cut-off etc~\citep[eg.,][]{Rucinski92,Rucinski07,Stepien06,Jiang12,Li19}. In addition, there also exists the Applegate mechanism~\citep{Apple92} and light-time effect (LTTE)~\citep[eg.,][]{Irwin59,Mayer90,Wolf99}, which can be used to explain the periodic changes of the orbital period of W UMa type stars. These physical characteristics make the contact binaries have a more complex evolutionary process. Accordingly, contact binaries are also of great significance for studying the formation and evolution of eclipsing binaries~\citep[eg.,][]{Qian17,Qian18}

ASAS J124343+1531.7 (2MASS J12434277+1531375, PY Com) was classified as an EW type binary star by~\citet{Gettel06} using the ROTSE-I telescope. Its period was reported as 0.266155 days, and the maximum and minimum V magnitude are 13.18 mag and 13.63 mag respectively. Later,~\citet{Hoffman09} defined ASAS J124343+1531.7 as a W UMa type star with a period of 0.30718 days. \citet{Drake14} obtained the period for 0.26616 days and the amplitude of variation for 0.4 mag. Minimal timings of ASAS J124343+1531.7 were obtained by~\citet{Diethelm09,Diethelm10,Diethelm11,Diethelm12}. Nevertheless, the study of photometric solutions and the period variations have not been published yet. This paper presents the complete $BVR_c$ light curves first obtained in 2018, 2019 and 2020. Thereafter, the $O-C$ analyses of orbital period and photometric solutions were put forward.

LINEAR 2323566 (2MASS J11514615+4750292) was first observed by the asteroid survey LINEAR~\citep{Palaversa13}, it was identified as an eclipsing binary with a period of 0.232872 days. The median LINEAR magnitude and amplitude variation were determined as 15.24 mag and 0.8 mag respectively. Later, \citet{Drake14} identified LINEAR 2323566 as an EW type binary star and determined a period of 0.23287 days by the Catalina Surveys Data Release-1 (CSDR1). It has no photometric analyses or period investigations. During our observations, two complete $BR_c$ light curves were obtained in 2016 and 2020 to do the photometric solution. Due to the lack of minima from other literatures, we only analyzed the orbital period using our observation data.

\section{Observations}
\subsection{Photometric Observations}
We observed ASAS J124343+1531.7 and LINEAR 2323566 using the 0.84-m f/15 Ritchey-Chr\'{e}tien telescope at Observatorio Astron\'{o}mico Nacional, Sierra San Pedro M\'{a}rtir (OAN-SPM), Baja California, M\'{e}xico, coupled with the Mexman filter-wheel and the Marconi3 CCD detector, equipped with Standard Johnson-Cousin-Bessel $UBVR_cI_c$ filters, in Feb., 2016, Apr., 2018 and Jan., 2019. As well as the WIYN 0.90m f/7.5 Cassegrain telescope at Kitt Peak National Observatory (KPNO), Arizona, USA, coupled with the Harris $UBVR_cI_c$ filter-wheel and the S2KB CCD detector followed up the observations in Feb. 2020. The log of observations is shown in Table 1. All these CCD images were reduced using the IRAF\footnote{IRAF is distributed by the National Optical Astronomy Observatories, which are operated by the Association of Universities for Research in Astronomy under cooperative agreement with the National Science Foundation.} package and has been corrected by bias and flat. In order to identify good comparison stars, and to determine the relative magnitudes of the variables, the field stars were calibrated during very photometric nights in which Landolt standard fields were also observed. Tables 2 and 3 list the HJD and the magnitudes of ASAS J124343+1531.7 and LINEAR 2323566. Figure 1 shows the light curves of the two objects in phase. The ephemeris used to convert time into phase is shown below,
\begin{equation}
\begin{split}
\label{equation:emphirical}
\textrm{ASAS J124343+1531.7: HJD} = Min.I_0 +0.266162 \times E, \\
\textrm{LINEAR 2323566: HJD} = Min.I_0 +0.232873 \times E,
\end{split}
\end{equation}
where the $Min.I_0$ is the primary minimum of light curve of each year. It can be seen from the light curves that both of them are W UMa type stars and show the O'Connell effect.

We also tried to find other sky survey data for these two targets, for example, All Sky Automated Survey (ASAS) \citep{Paczynski06}, All-Sky Automated Survey for Supernovae (ASAS-SN) \citep{Shappee14}, Catalina Sky Surveys (CSS) \citep{Drake14}, TESS \citep{Ricker15}, Northern Sky Variability Survey (NSVS) \citep{Wozniak04}, etc. Although these two targets were both observed, the quality of survey data except TESS is not good enough to support subsequent analysis work. Finally, we adopted only the TESS data for ASAS J124343+1531.7. The TESS data processed by $lightkurve$ package \citep{LK18} were downloaded from the Mikulski Archive for Space Telescopes (MAST)\footnote{https://mast.stsci.edu/portal/Mashup/Clients/Mast/Portal.html}. The time span is about 25 days from March 21, 2020 to April 15, 2020. In order to better show the overall light curve, we used Equation (1) to convert the BJD of TESS data into phase and obtained the complete light curves in one period plotted in Figure 2. Later, we also used this data for orbit period and photometric analysis.

\begin{table}
\begin{center}
\scriptsize
\caption{Information of observation for ASAS J124343+1531.7 and LINEAR 2323566}
\begin{tabular}{lcccc}
\hline\hline
Objects	& RA DEC & Date	& Filter(Exposure Time [s]) & Filter(Number of data points)  \\
\hline
\multirow{3}{*}{ASAS J124343+1531.7} & \multirow{3}{*}{12 43 42.77 +15 31 37.46}
  & Apr. 22, 2018  & \multirow{3}{*}{$B(50), V(25), R_c(15)$} & $B(63),V(63),R_c(63)$ \\
 & & Jan. 24, 2019  & & $B(86),V(87),R_c(87)$  \\
 & & Feb. 15, 2020  & & $B(42),V(42),R_c(42)$   \\
\multirow{2}{*}{LINEAR 2323566} & \multirow{2}{*}{11 51 46.16 +47 50 29.45}
  & Mar. 20, 2016  & \multirow{2}{*}{$B(50\&100), R_c(30)$} & $B(263),R_c(131)$       \\
&  & Feb. 16, 2020  &  & $B(78),R_c(39)$   \\
\hline
\end{tabular}
\end{center}
\end{table}

\begin{table}[htbp]
  \centering
  \caption{Photometric data of ASAS J124343+1531.7}
    \begin{tabular}{cccccccc}
    \hline
    \multicolumn{2}{c}{B} &       & \multicolumn{2}{c}{V} &       & \multicolumn{2}{c}{R} \\
    \cline{1-2} \cline{4-5} \cline{7-8}
    HJD &  M(mag)  &                 & HJD  &  M(mag) &                 & HJD   &  M(mag) \\
    \hline
    \multicolumn{8}{c}{April 22, 2018}\\
    2458230.69130  & 14.287  &       & 2458230.69072  & 13.417  &       & 2458230.69031  & 12.976  \\
    2458230.69337  & 14.316  &       & 2458230.69279  & 13.438  &       & 2458230.69238  & 13.004  \\
    2458230.69544  & 14.372  &       & 2458230.69486  & 13.469  &       & 2458230.69445  & 13.026  \\
    2458230.69751  & 14.403  &       & 2458230.69693  & 13.510  &       & 2458230.69652  & 13.058  \\
    2458230.69958  & 14.449  &       & 2458230.69900  & 13.561  &       & 2458230.69859  & 13.087  \\
    ... & ... &                      & ... & ...  &                     & ... & ... \\
    2458895.02813  & 14.260  &       & 2458895.02665  & 13.372  &       & 2458895.02518  & 12.934  \\
    2458895.03320  & 14.363  &       & 2458895.03136  & 13.433  &       & 2458895.02982  & 12.985  \\
    2458895.03834  & 14.462  &       & 2458895.03626  & 13.527  &       & 2458895.03471  & 13.070  \\
    2458895.04435  & 14.585  &       & 2458895.04208  & 13.621  &       & 2458895.04039  & 13.162  \\
    2458895.05094  & 14.583  &       & 2458895.04856  & 13.674  &       & 2458895.04703  & 13.232  \\
    \hline
    \end{tabular}%
  \label{tab:addlabel}%
\end{table}%

\begin{table}[htbp]
  \centering
  \caption{Photometric data of LINEAR 2323566}
    \begin{tabular}{ccccc}
    \hline
    \multicolumn{2}{c}{B} &       & \multicolumn{2}{c}{R} \\
    \cline{1-2} \cline{4-5}
     HJD  &  M(mag) &                 & HJD   &  M(mag) \\
     \hline
    \multicolumn{5}{c}{March 20, 2016}\\
    2457467.65545  & 16.732  &       & 2457467.65651  & 15.021  \\
    2457467.65731  & 16.710  &       & 2457467.65943  & 15.020  \\
    2457467.65835  & 16.706  &       & 2457467.66236  & 14.992  \\
    2457467.66023  & 16.730  &       & 2457467.66528  & 14.979  \\
    2457467.66127  & 16.683  &       & 2457467.66820  & 14.951  \\
    ...  &  ... &                    & ...  &  ...  \\
    2458896.03358  & 17.004  &       & -      & - \\
    2458896.03710  & 17.104  &       & -      & - \\
    2458896.03902  & 17.128  &       & -      & - \\
    2458896.04271  & 17.221  &       & -      & - \\
    2458896.04459  & 17.307  &       & -      & - \\
    \hline
    \end{tabular}%
  \label{tab:addlabel}%
\end{table}%

\begin{figure}\centering
\includegraphics[width=0.6\textwidth]{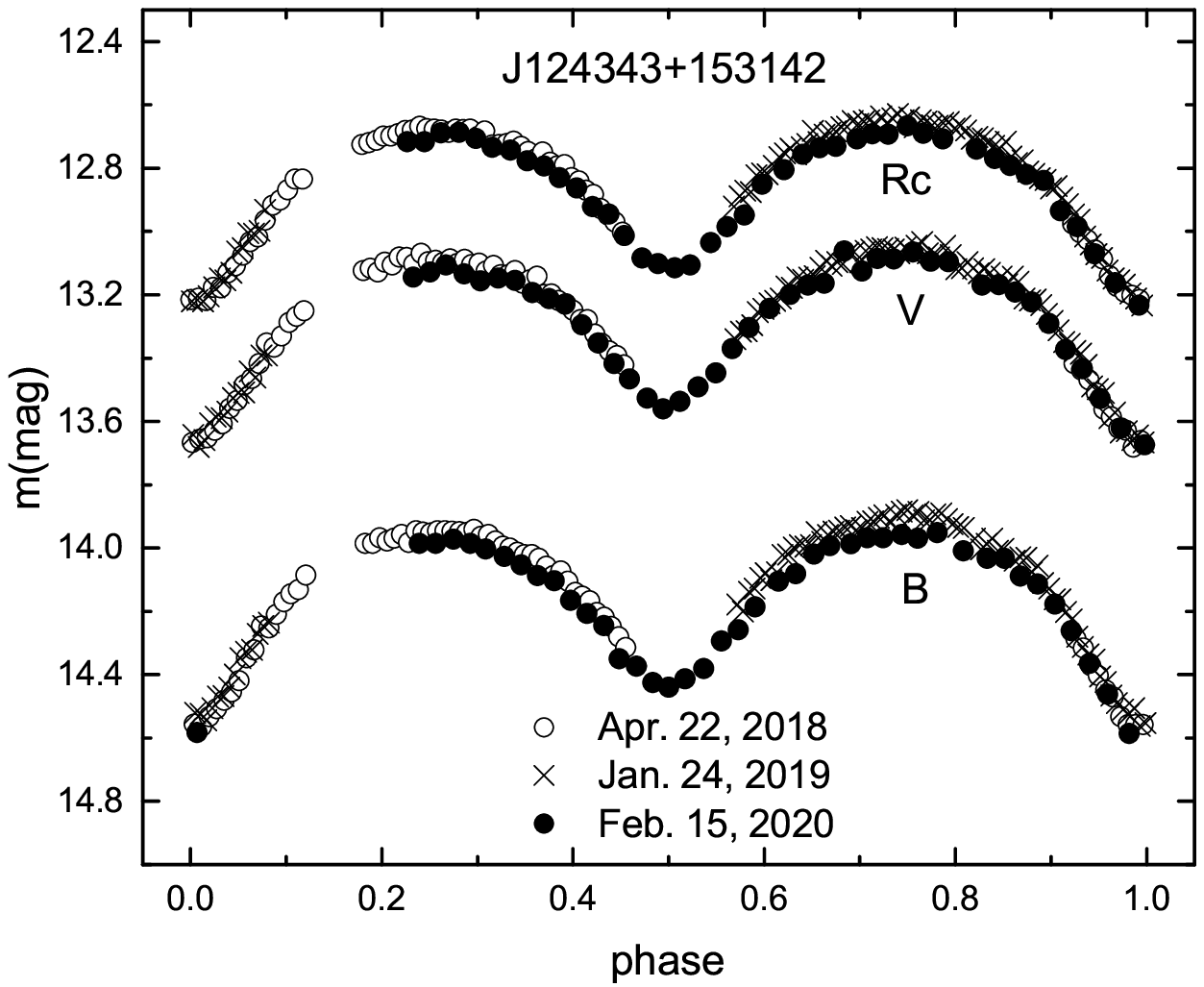}\\
\includegraphics[width=0.6\textwidth]{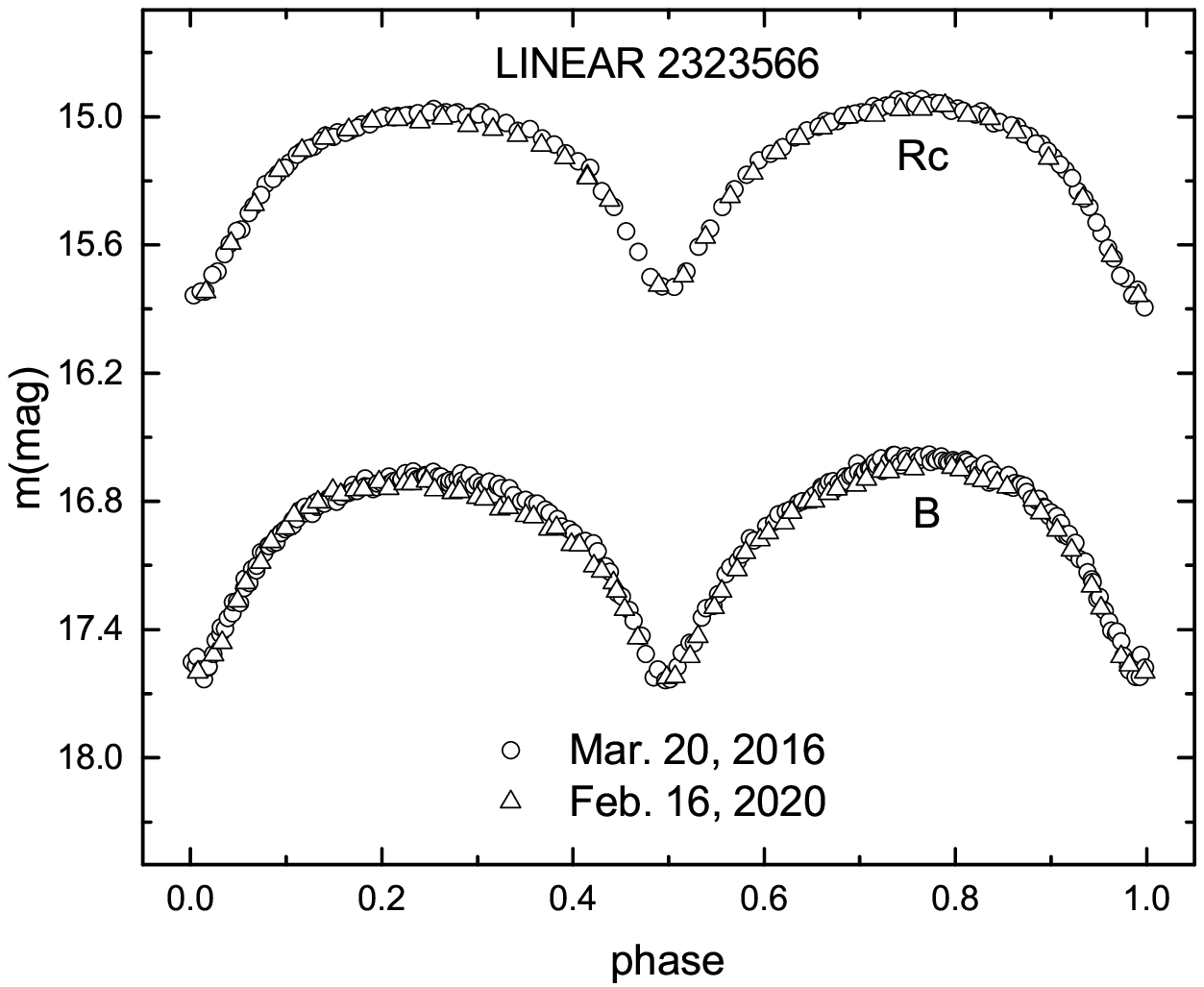}
\caption{The observed light curves of ASAS J124343+1531.7 and LINEAR 2323566.}
\end{figure}

\begin{figure}\centering
\includegraphics[width=0.6\textwidth]{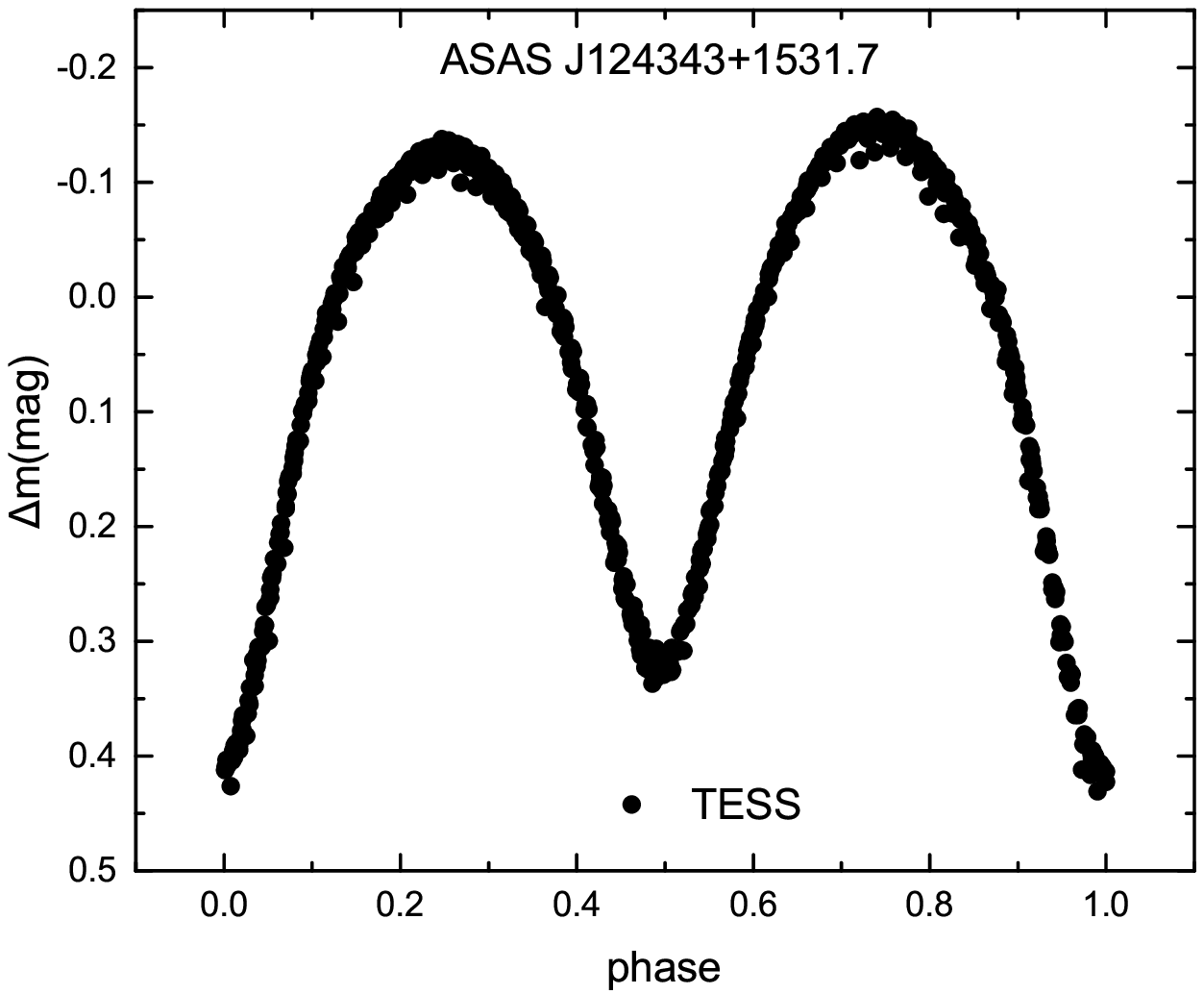}
\caption{The TESS light curves of ASAS J124343+1531.7 observed from March 21, 2020 to April 15, 2020 and represented by phase.}
\end{figure}

\subsection{Spectroscopic Observations of LAMOST}
Guoshoujing Telescope (the Large Sky Area Multi-Object Fiber Spectroscopic Telescope (LAMOST)) is a 4m reflecting Schmidt telescope equipped with 4000 fibers with a 5$^{\circ}$ field of view \citep{Cui12,Zhao12}. The wavelength coverage ranges from 3700 \AA ~to 9000 \AA ~\citep{Du16}, and the overall spectral resolution of LAMOST is approximately 1800. The data were reduced with LAMOST pipelines \citep{Luo12}. And the LAMOST spectral analysis pipeline (also called 1D pipeline, LASP) is used to classify and measure the spectra \citep{Luo01,Luo04,Luo08,Luo12,Wang10}. Through LASP, the observed stellar spectra are classified into different subclasses by matching with template spectra \citep{Wei14}. The spectra and parameters of ASAS J124343+1531.7 and LINEAR 2323566 were obtained from Data Release 7\footnote{http://dr7.lamost.org/} \citep{Luo15}. Table 4 lists the spectral information of these two targets, including the observational data, Heliocentric Julian date,  phase, spectral type, effective temperature ($T_{eff}$), $log(g)$, radial velocity (RV) and signal to noise ratio of r filter (SNR-r). We downloaded these spectra and plotted them in Figure 3.

\begin{table}
\begin{center}
\scriptsize
\caption{The spectral information of ASAS J124343+1531.7 and LINEAR 2323566 released by LAMOST}
\setlength{\leftskip}{-55pt}
\begin{tabular}{lccccccccc}
\hline
Targets & Obsdate & HJD(d)  & Phase & Subclass & $T_{eff}(K)$ & $log(g)$ & RV (km s$^{-1}$) & SNR-r &  EWs: H$_\alpha$ (\AA)\\
\hline
ASAS J124343+1531.7 & Mar. 9, 2014  & 2456726.12748 & 0.11390 & G9 & 5221.76$\pm$50.97 & 4.342 & -24.29  & 93.2  &  3.019$\pm0.018$   \\
                    & Dec. 18, 2017 & 2458106.42784 & 0.05481 & G9 & 5083.87$\pm$27.15 & 4.230 & -9.46   & 119.8 &  3.361$\pm0.033$   \\
LINEAR 2323566      & Jan. 21, 2014 & 2456679.31742 & 0.39520 & K5 & 4464.32$\pm$88.88 & 4.428 & 27.85   & 48.2  &  4.663$\pm0.083$    \\
\hline
\end{tabular}
\end{center}
\end{table}

\begin{figure}\centering
\includegraphics[width=0.47\textwidth]{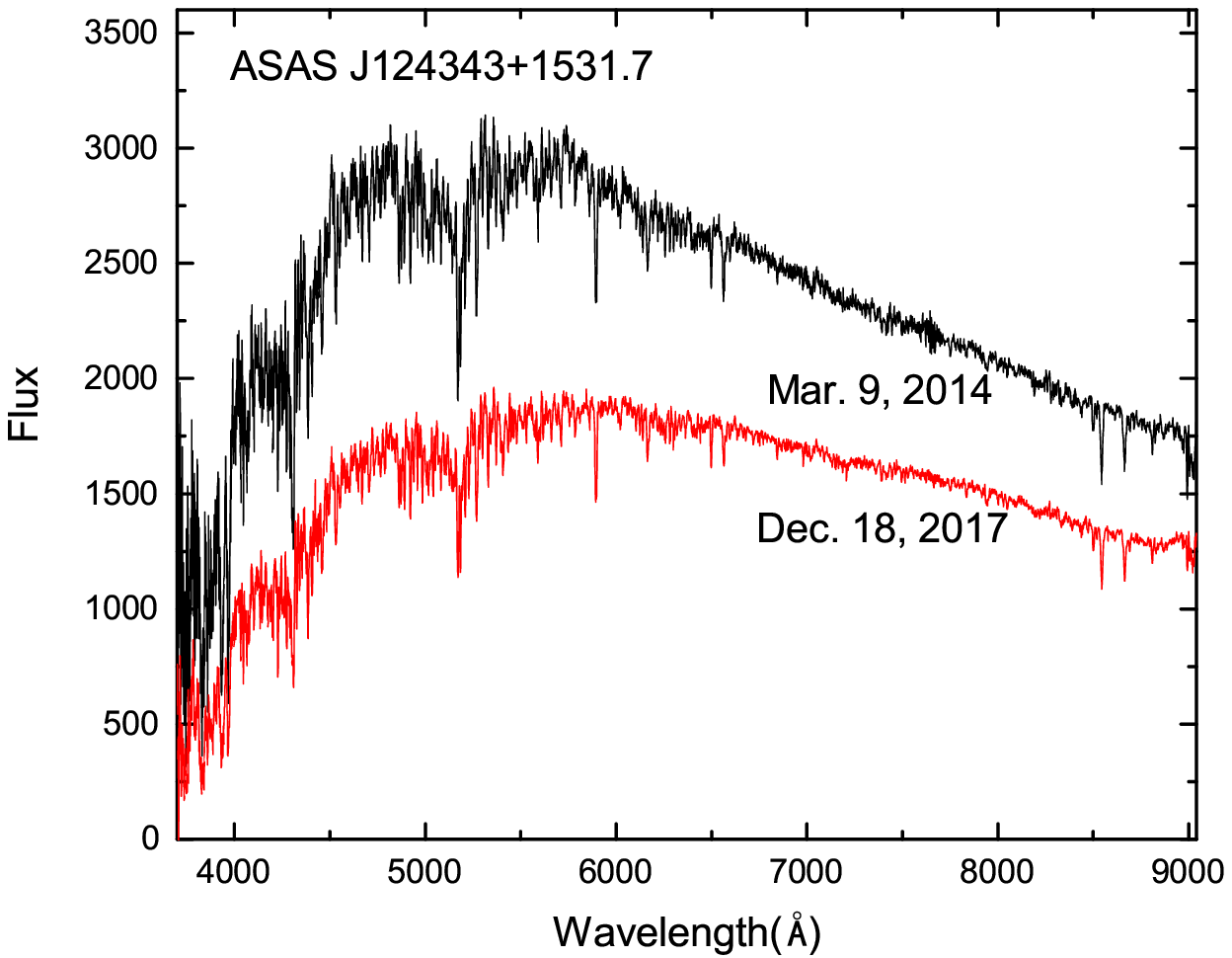}
\includegraphics[width=0.47\textwidth]{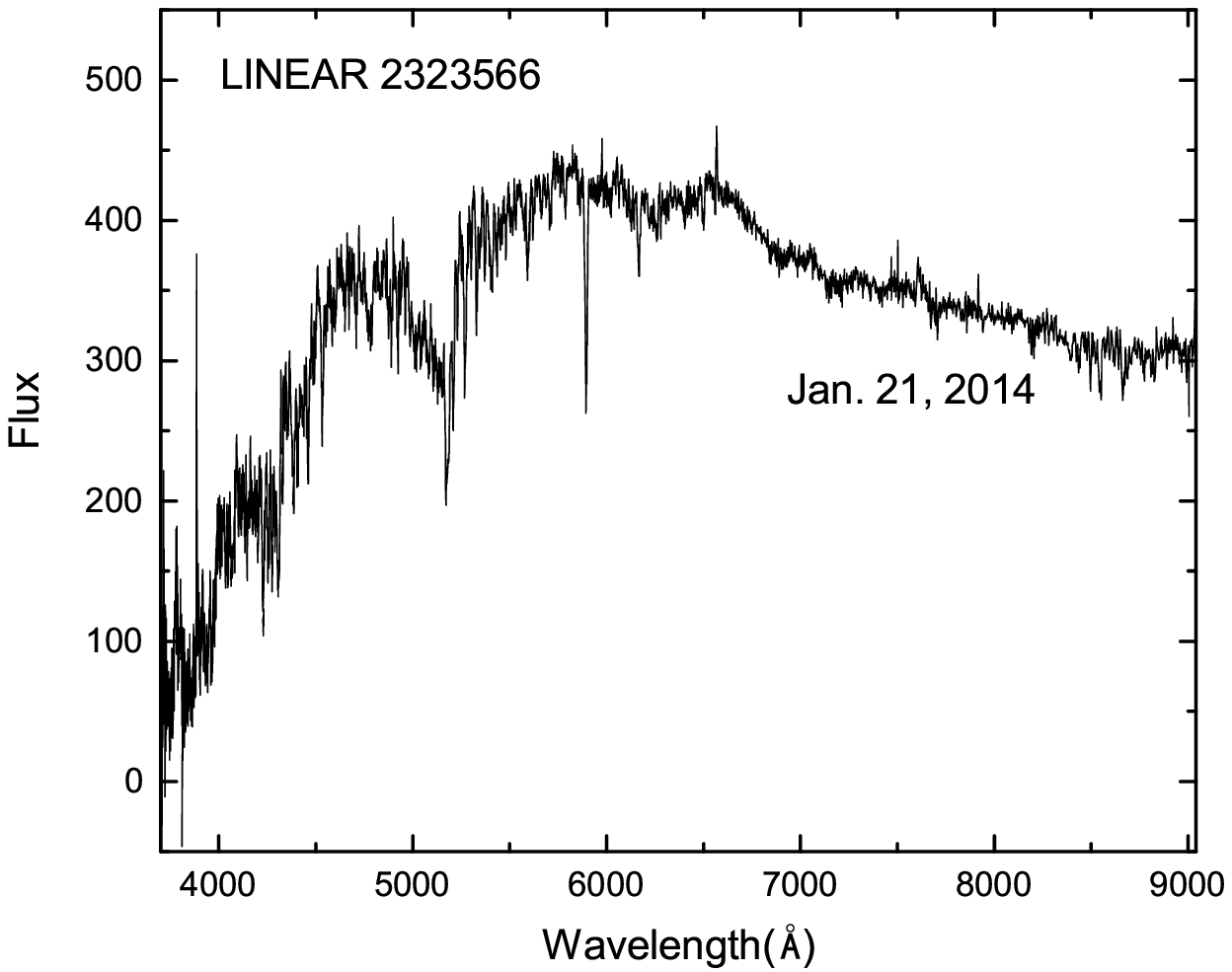}
\caption{Two spectra of ASAS J124343+1531.7 (left panel) and one spectrum of LINEAR 2323566 (right panel).}
\end{figure}

\section{Photometric Solutions}
The 2015 version of the Wilson-Devinney (W-D) \citep{Wilson71,Wilson79,Wilson90,Wilson07} program was used to analyze the new multi-passband CCD photometry and obtained the model parameters. The effective temperatures used in the W-D program are from LAMOST listed in Table 4. Since ASAS J124343+1531.7 has two spectra, we assigned an average temperature of 5153$\pm$153 K as the final effective temperature of the primary star. The effective temperature of the primary star used in calculation was 4464$\pm$89 K for LINEAR 2323566. According to \citet{Lucy67} and \citet{Rucinski69}, the gravity-darkening coefficients and bolometric albedo coefficients were set as $g_{1,2}$ = 0.32 and $A_{1,2}$ = 0.5, respectively. Bolometric limb-darkening and bandpass limb-darkening coefficients were obtained from \citet{van93}. The $TESS$ band is added to the limb-darkening table by Professor Van Hamme (private communication with Dr. Wang, Z. H.). During our modeling, Mode 3 was chosen for the two binary systems. Other parameters include the orbital inclination $i$, the effective surface temperature of the secondary star $T_2$, the dimensionless potentials of two components $\Omega_1=\Omega_2$ and the monochromatic luminosity of the primary star $L_1$ were set as adjustable parameters.

First, we used all the observed data from OAN-SPM, KPNO, and TESS for ASAS J124343+1531.7 and LINEAR 2323566 to do the $q$-search process respectively, to determine their mass ratios. Then, we obtained solutions with mass ratios ranging from 0.1 to 10 with an interval of 0.1. The relationship between their mass ratios and residuals is shown in Figure 4. The mass ratios corresponding to their minimum residuals are 3.6 and 1.4, respectively. Consequently, we set them as the initial values of $q$ and adjustable, and thereafter, the convergent solutions for mass ratios 3.758 and 1.438 were obtained. The photometric solutions corresponding to these mass ratios are considered to be the final solution of these two objects, and are applied to determine the physical parameters and discuss the evolutionary status. All the parameters and errors of photometric solutions are listed in Table 5.

Since the two maxima of the light curves are not equal for these two objects, this indicates that they both exist O'Connell effect, which is generally caused by the existence of spots~\citep[eg., V789 Her and V1007 Cas,][]{Li18b}. Therefore, we added spot parameters when using W-D program and fitted them according to observations at different times to study the changes of spots. We divide the data of ASAS J124343+1531.7 into three parts, 2018 \& 2019, Feb., 2020 for our observations and Mar. \& Apr., 2020 for TESS, while the data of LINEAR 2323566 into two parts, 2016 and 2020 for our observations. Their mass ratios are respectively set to 3.758 and 1.438, then the spot parameters are adjusted to obtain the best fitting solution. We have tried adding cool spot or hot spot on the primary or secondary stars, and finally we obtained the best results when we used a single cool spot on the hotter star. The theoretical light curves of different times are plotted in Figure 5, it can be seen that the two maxima of light curves fit well, the residuals between observed data and theoretical light curves are almost flat. The spot parameters at different times are listed in Table 6.

The geometric structure under different spot parameters at phase 0.25 and 0.75 are displayed in Figure 6, where the model of the binary systems and the location of the cool spot are displayed more intuitive. The dimension parameters of the corresponding geometric structure are also listed in Table 6. The errors obtained by the W-D method are mathematical fitting and have no physical meaning.

\begin{table*}
\begin{center}
\caption{Physical parameters of ASAS J124343+1531.7 and LINEAR 2323566}
\begin{tabular}{lcc}
\hline\hline
Star  &  ASAS J124343+1531.7   & LINEAR 2323566      \\
\hline
$T_1(K)$            &   5153             &    4467                 \\
$q(M_2/M_1) $       &   3.758 $\pm$0.028        &    1.438 $\pm$0.040     \\
$T_2$               &   4833$\pm$5              &    4382$\pm$10          \\
$i(deg)$            &   77.4$\pm$0.2            &    87.4 $\pm$0.6        \\
$\Omega_1=\Omega_2$ &   7.404$\pm$0.037         &    4.348 $\pm$0.063     \\
$L_{1B}/L_{B}$      &   0.335$\pm$0.003         &    0.465$\pm$ 0.003    \\
$L_{1V}/L_{V}$      &   0.314$\pm$0.003         &    -                    \\
$L_{1R_c}/L_{R_c}$  &   0.299$\pm$0.003         &    0.449$\pm$ 0.003    \\
$L_{1TESS}/L_{TESS}$&   0.290$\pm$0.001         &    -                   \\
$r_1$               &   0.287 $\pm$0.001        &    0.358$\pm$0.003      \\
$r_2$               &   0.515 $\pm$0.004        &    0.423$\pm$0.013       \\
$f$                 &   31.8$\pm$ 5.8 \%        &    14.9$\pm$10.9 \%      \\
\hline
\end{tabular}
\end{center}
\end{table*}

\begin{figure}\centering
\includegraphics[width=0.48\textwidth]{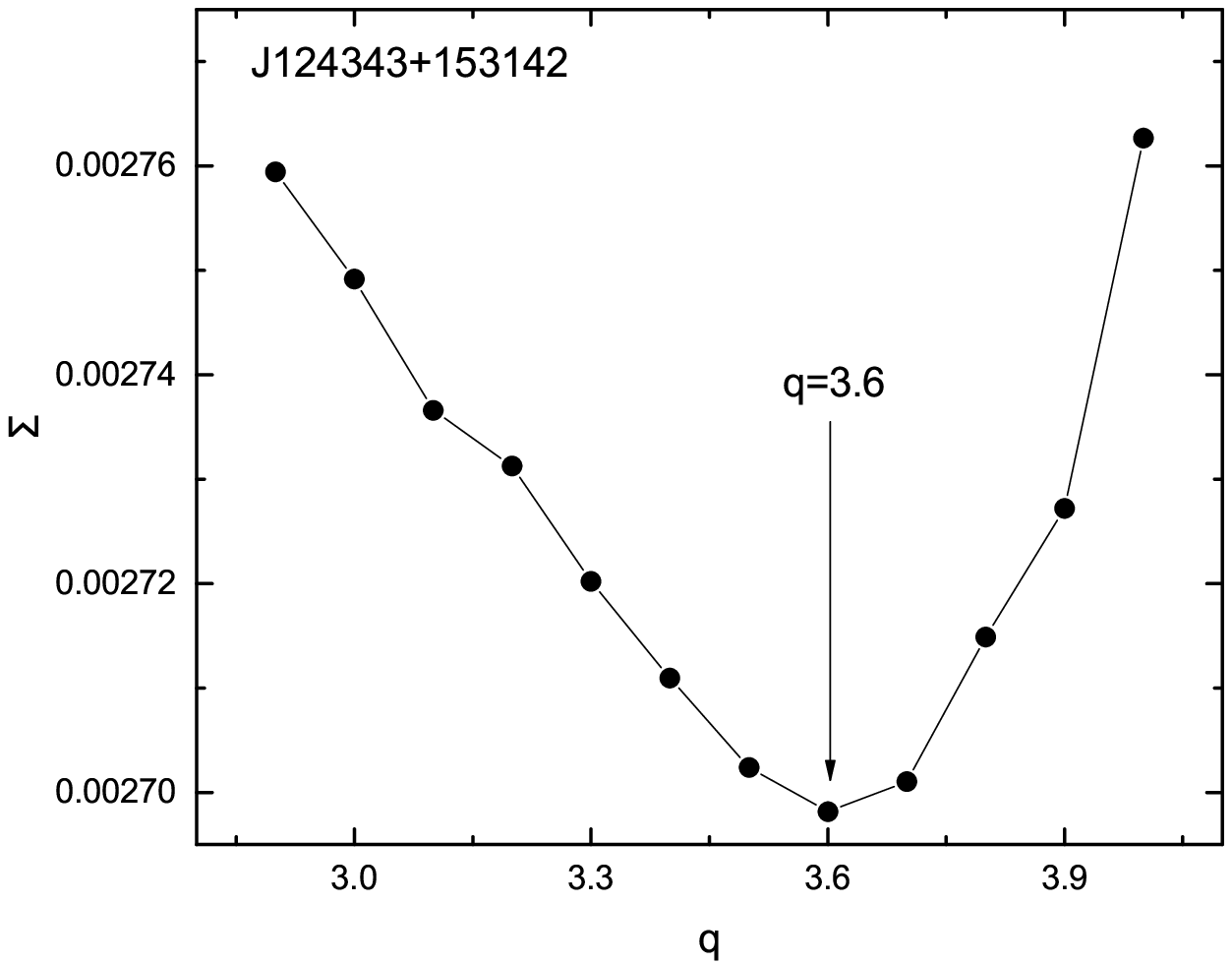}
\includegraphics[width=0.48\textwidth]{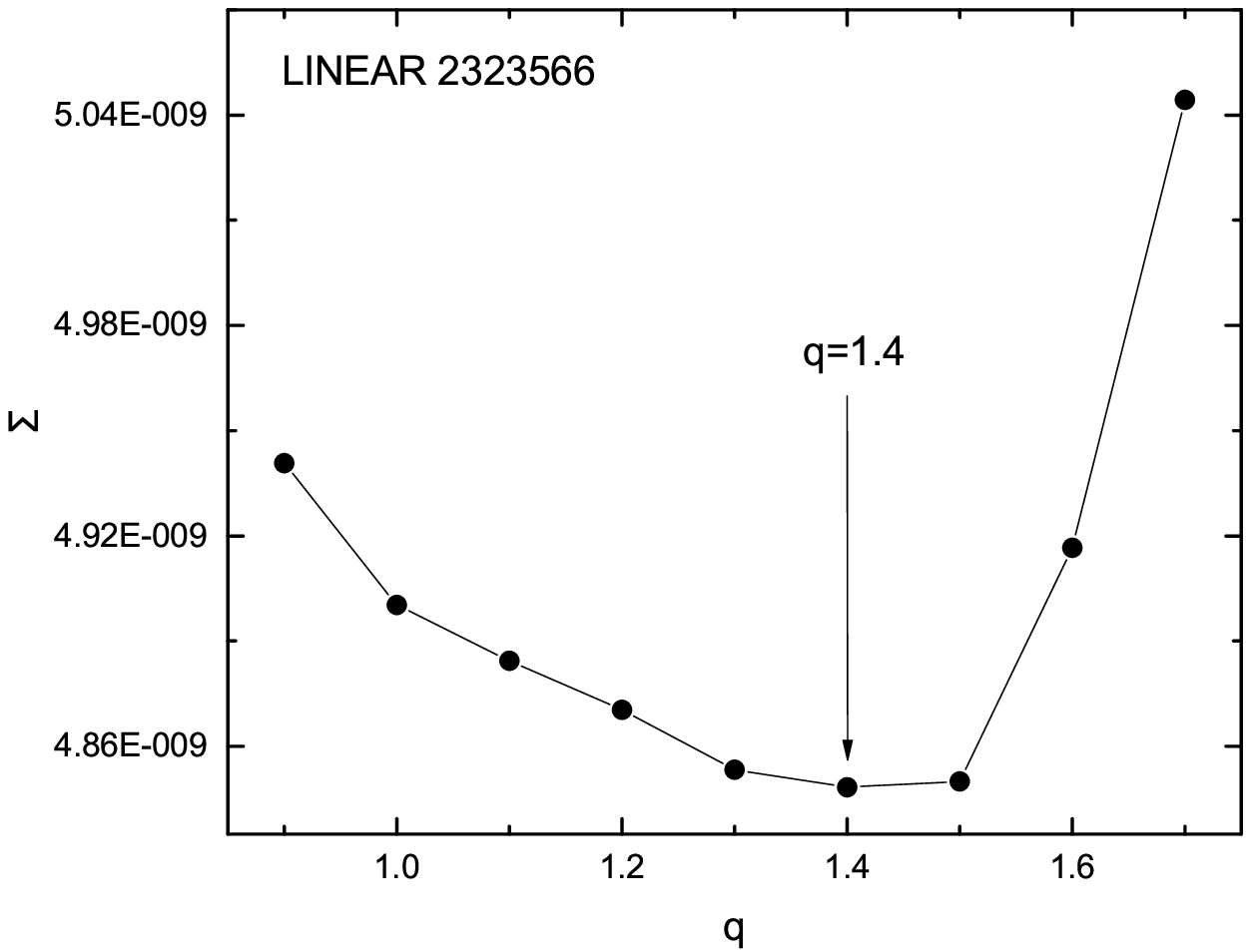}
\caption{$\Sigma$ - q curves of ASAS J124343+1531.7 (left panel) and LINEAR 2323566 (right panel).}
\end{figure}

\begin{figure}\centering
\includegraphics[width=0.60\textwidth]{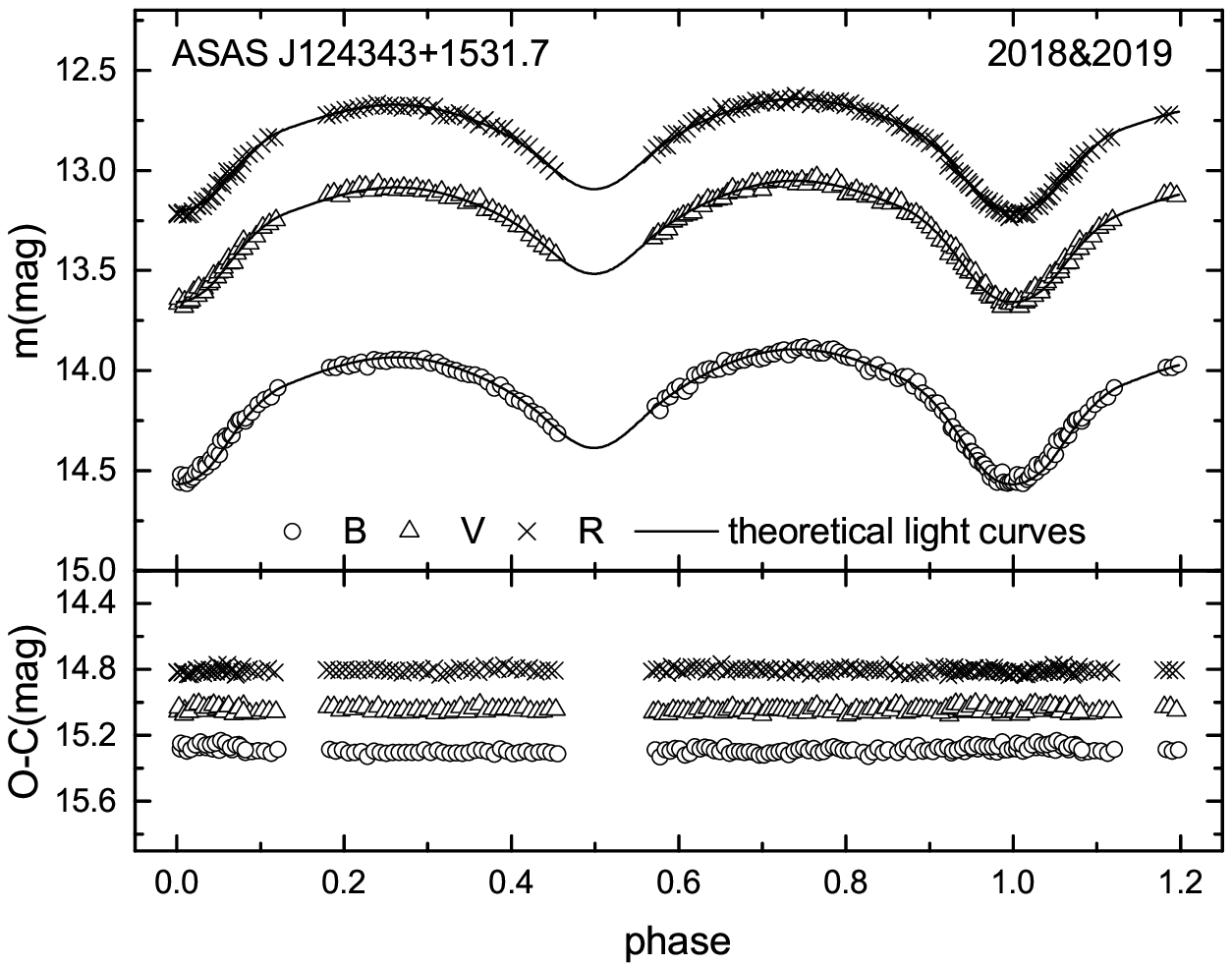}\\
\includegraphics[width=0.60\textwidth]{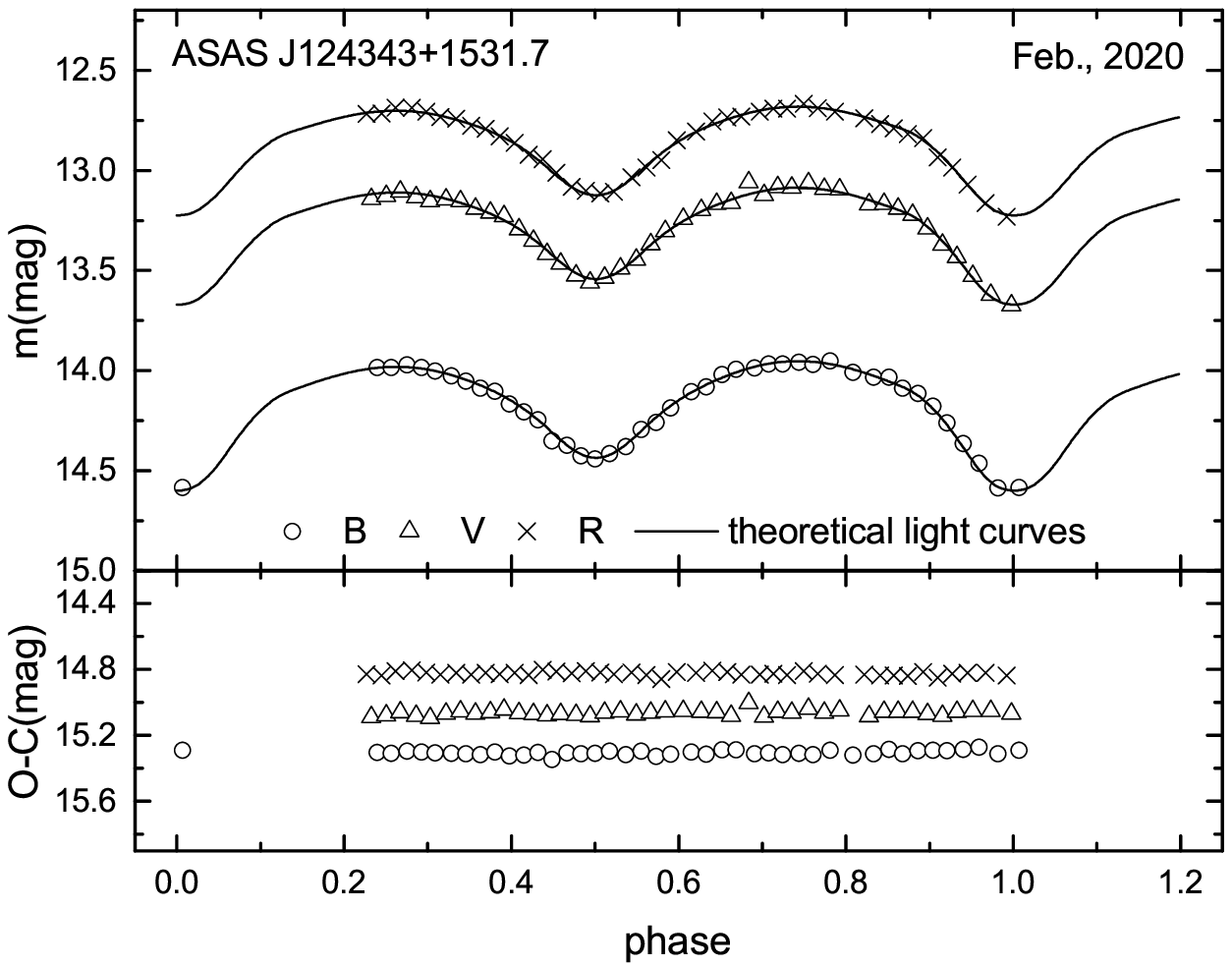}\\
\includegraphics[width=0.60\textwidth]{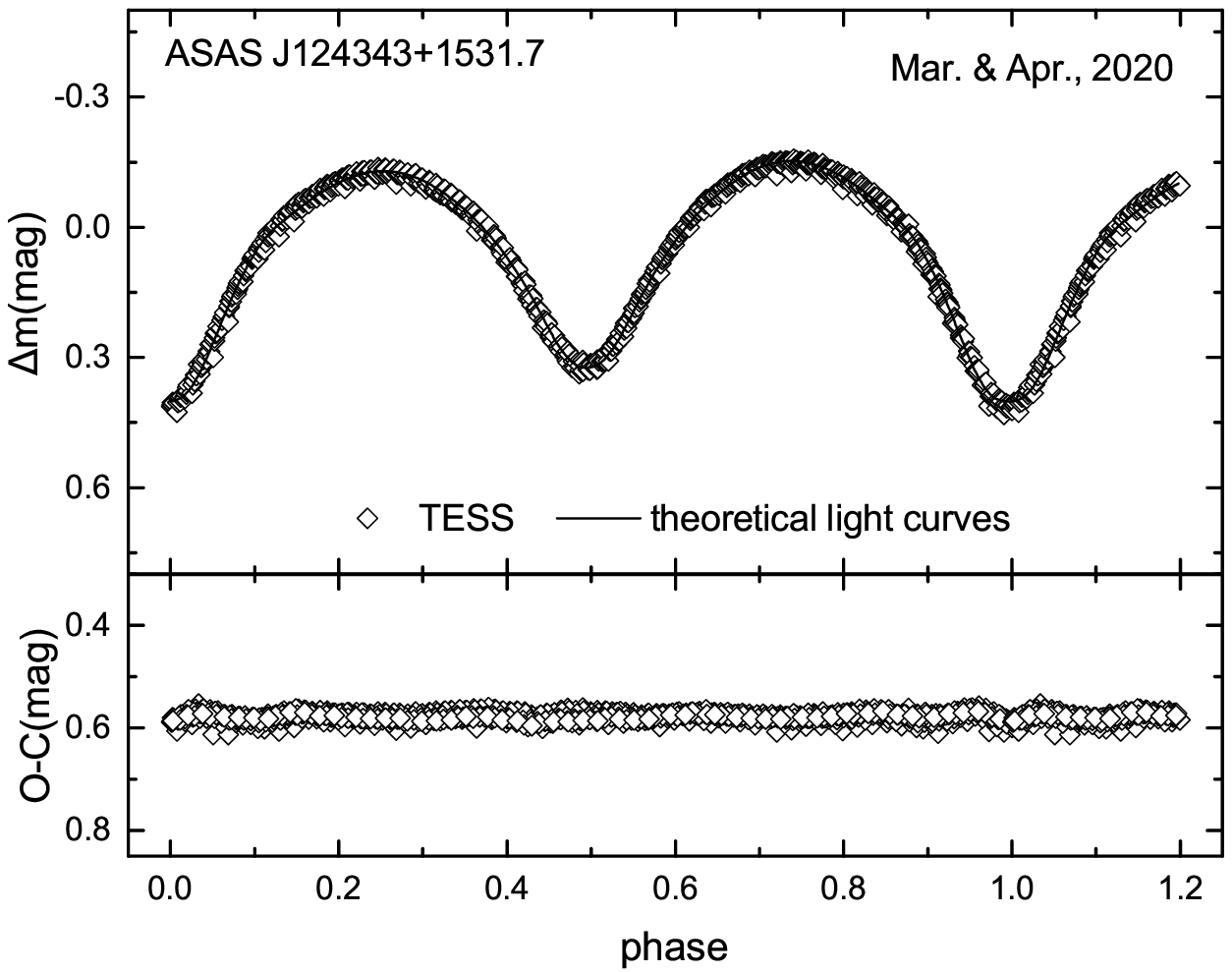}
\caption{Theoretical light curves (solid line) fitted by WD code of ASAS J124343+1531.7 and LINEAR 2323566. They all employed a cool spot on their primary stars.}
\end{figure}

\begin{figure}\centering
\includegraphics[width=0.60\textwidth]{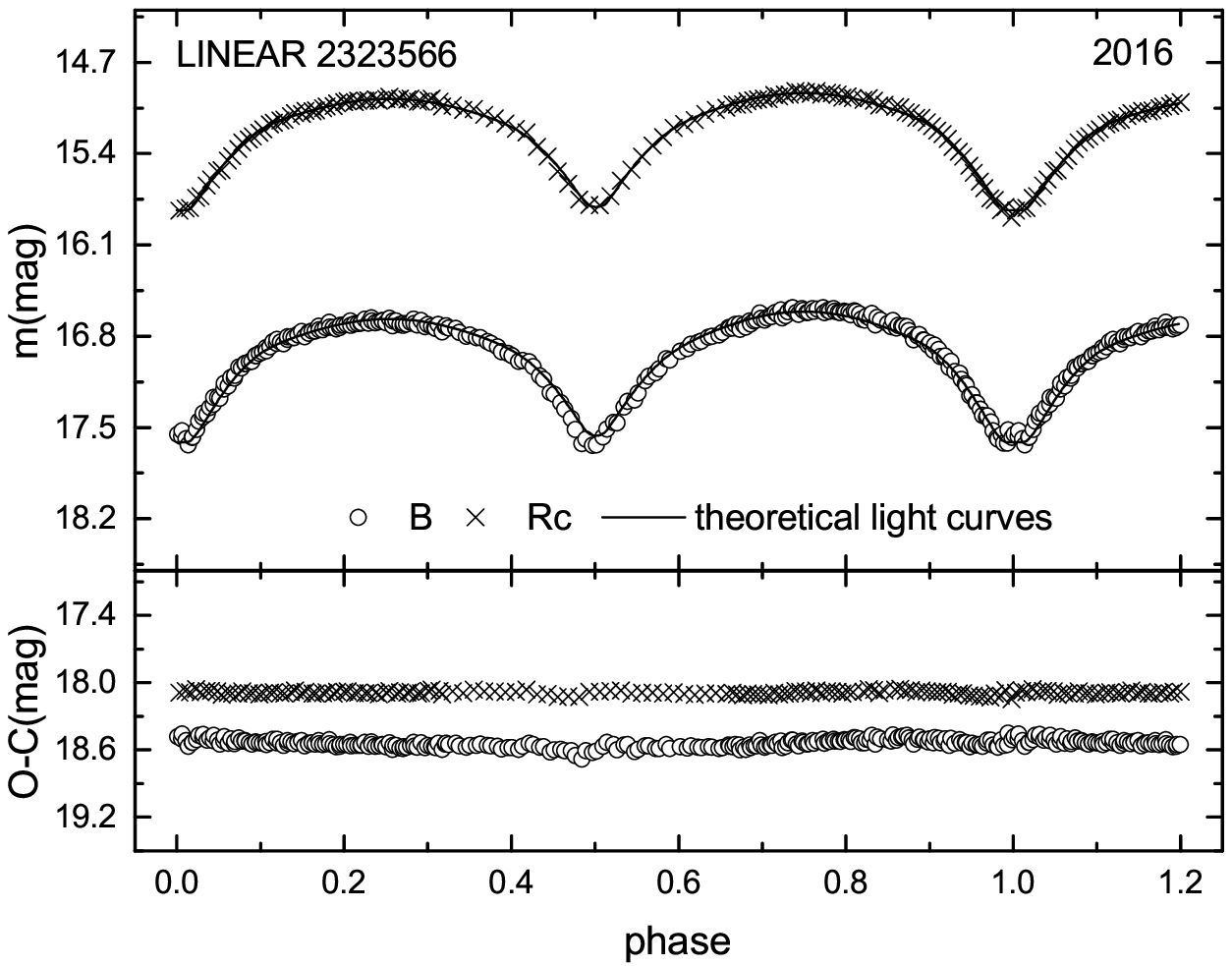}\\
\includegraphics[width=0.60\textwidth]{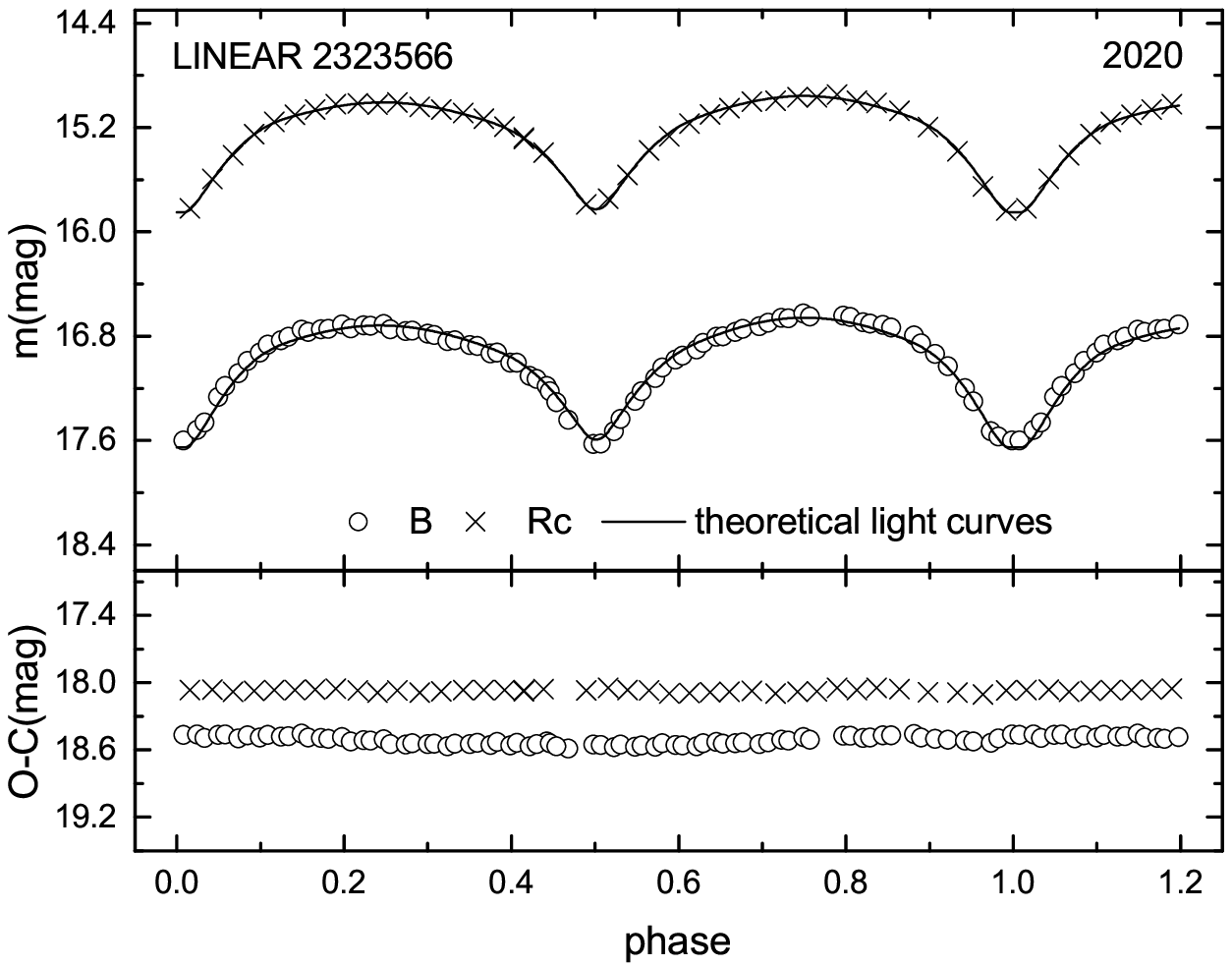}\\
\textbf{Fig. 5 - Continued}
\end{figure}

\begin{table*}
\begin{center}
\caption{Spot parameters and geometric structure of ASAS J124343+1531.7 and LINEAR 2323566}
\setlength{\leftskip}{-10pt}
\begin{tabular}{lccccc}
\hline
Star  &  \multicolumn{3}{c}{ASAS J124343+1531.7}   & \multicolumn{2}{c}{LINEAR 2323566}      \\
Year  &  2018 \& 2019  & Feb., 2020  &  Mar. \& Apr., 2020  & 2016  &  2020  \\
\hline
Spot                &   Star 1                 &   Star 1                   &    Star 1              &     Star 1  &     Star 1         \\
$\theta(deg)$       &  86.9    $\pm$9.1        &  105    $\pm$8      &  102    $\pm$2               &  101    $\pm$27       &   103  $\pm$14   \\
$\lambda(deg)$      &  275.3   $\pm$4.1        &  275    $\pm$14     &  257    $\pm$2               &  269    $\pm$3        &   249  $\pm$4     \\
$r_s(deg)$          &  17.0    $\pm$0.5        &  16     $\pm$1      &  17     $\pm$1               &  17     $\pm$0        &   19   $\pm$1    \\
$T_s$               &  0.71    $\pm$0.03       &  0.71   $\pm$0.08   &  0.71   $\pm$0.01            &  0.73   $\pm$0.03      &    0.72  $\pm$0.05    \\
$r_1(pole)$         &  0.261   $\pm 0.001 $    &  0.258  $\pm0.001 $& 0.267  $\pm 0.000 $         &  0.338  $\pm 0.001 $   &  0.334  $\pm 0.001 $   \\
$r_1(side)$         &  0.273   $\pm 0.001 $    &  0.270  $\pm0.002 $& 0.280  $\pm 0.000 $         &  0.355  $\pm 0.001 $   &  0.351  $\pm 0.001 $   \\
$r_1(back)$         &  0.314   $\pm 0.002 $    &  0.307  $\pm0.003 $& 0.327  $\pm 0.001 $         &  0.394  $\pm 0.001 $   &  0.388  $\pm 0.002 $   \\
$r_2(pole)$         &  0.474   $\pm 0.001 $    &  0.471  $\pm0.001 $& 0.479  $\pm 0.000 $         &  0.398  $\pm 0.001 $   &  0.394  $\pm 0.001 $   \\
$r_2(side)$         &  0.513   $\pm 0.001 $    &  0.509  $\pm0.002 $& 0.521  $\pm 0.000 $         &  0.423  $\pm 0.001 $   &  0.418  $\pm 0.001 $   \\
$r_2(back)$         &  0.541   $\pm 0.001 $    &  0.536  $\pm0.002 $& 0.550  $\pm 0.001 $         &  0.457  $\pm 0.001 $   &  0.451  $\pm 0.002 $   \\
\hline
\end{tabular}
\end{center}
\end{table*}

\begin{figure}\centering
\includegraphics[width=0.80\textwidth]{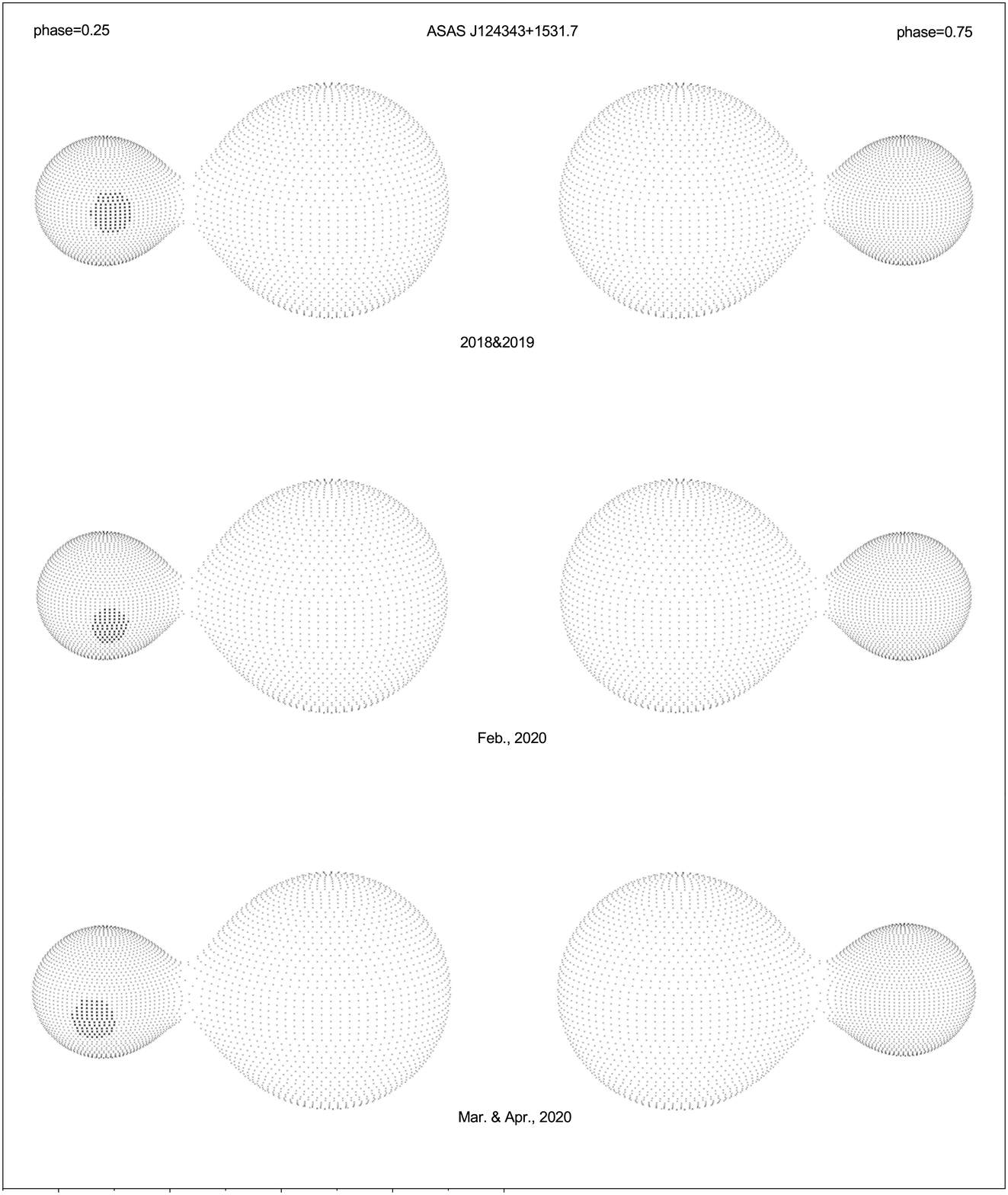}\\
\caption{Geometrical configuration of ASAS J124343+1531.7 and LINEAR 2323566 at phase 0.25 and 0.75. For both binaries, the smaller component is designated as the primary star, because it is the hotter star, although it is less massive.}
\end{figure}

\begin{figure}\centering
\includegraphics[width=0.80\textwidth]{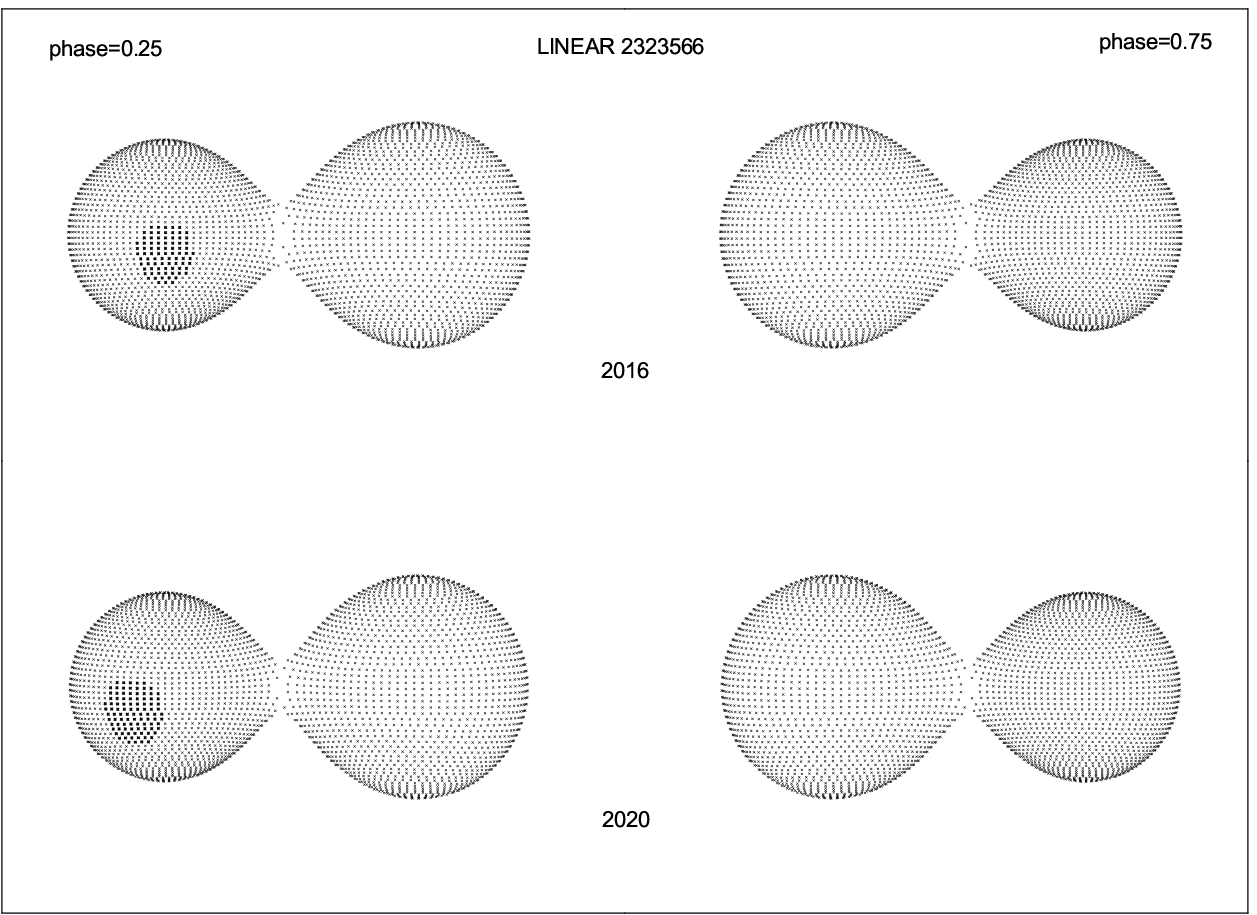}\\
\textbf{Fig. 6 - Continued}
\end{figure}

\section{Orbital Period Analysis}
In order to analyze the orbital period changes of ASAS J124343+1531.7 and LINEAR 2323566, we used the K-W method \citep{Kwee56} to calculate the eclipsing times. However, the time resolution of TESS data for ASAS J124343+1531.7 is low, on account of the 30 minute observation cadence. Therefore, we used the phase shift method to apply higher density light curves. First, we divided TESS light curves into four parts, the BJD time span of these four parts are 2458929.51622 to 2458935.24541, 2458936.26624 to 2458940.82870, 2458946.74526 to 2458951.32849 and 2458951.34932 to 2458954.80755, and the number of points in each part is 234, 215, 165 and 166, respectively. Second, using the equation BJD = BJD$_0$ + P $\times$ E, where BJD is the observing time, BJD$_0$ is the reference time for each part, which are 2458932.120404, 2458938.245397, 2458948.620222 and 2458952.620118, and P is the orbital period of ASAS J124343+1531.7. Ultimately, we moved the long time-span data into one period and determined the four panels of Figure 8. Then, we obtained 8 eclipsing times from these four sets of light curves.

Combining other literatures, public data and our own observations, we have collected 16 and 4 minima for ASAS J124343+1531.7 and LINEAR 2323566 respectively shown in Table 7. The ephemerides used to calculate $O-C$ (observation-calculation) values are:
\begin{eqnarray}
\begin{split}
\textrm{ASAS J124343+1531.7: BJD} = 2458230.71179 + 0.266162 \times E, \\
\textrm{LINEAR 2323566: BJD} = 2457467.73550 + 0.2328734 \times E.
\end{split}
\end{eqnarray}
Since the minimum moments of TESS data for ASAS J124343+1531.7 are given in BJD, we converted other observed minima from HJD to BJD.  The calculated epoch and $O-C$ values were also listed in Table 7, and the corresponding data points were plotted in Figure 8.

The trend of ASAS J124343+1531.7 shows slightly nonlinear variations. Therefore we fitted the $O-C$ values with an upward parabola, where the revised ephemeris is as follows,
\begin{eqnarray}
\begin{split}
\textrm{Min.I} = 2458230.71144(\pm0.00036) + 0.2661615(\pm0.0000001) \times E\\
 + 4.517(\pm1.329) \times 10^{-11} \times E^2.
\end{split}
\end{eqnarray}
From Equation (3), the coefficients of second-order term are positive, indicating that the period of ASAS J124343+1531.7 is secular increase and the rate is calculated as $dp/dt$ = 1.239 $\times$ 10$^{-7}$ $d\cdot yr^{-1}$. After removing the parabolic term from $O-C$ trend, the residuals are plotted in the lower panel. The residuals are almost flat, which indicates the $O-C$ values were fitted well. While LINEAR 2323566 has very few data points, and the distribution of data points shows no nonlinear change, as shown on Figure 8. Therefore, we only performed simple linear fitting on LINEAR 2323566, and obtained the new ephemeris as follows,
\begin{eqnarray}
\begin{split}
\textrm{Min.I} = 2457467.73527(\pm0.00013) + 0.23287207(\pm0.00000007) \times E.
\end{split}
\end{eqnarray}
The residuals of the fitting are also displayed in the bottom panel of Figure 8, and the corresponding data are listed in Table 7.

Due to the 2015 version of the W-D program can also analyze the times of minima, then we included the minima data of these two objects during the W-D process to solve the epoch, period and orbital period variation. Then, we transformed the results into the form of Equations (3) and (4) as shown below,
\begin{eqnarray}
\begin{split}
\textrm{Min.I} = 2458230.71143(\pm0.00011) + 0.2661615(\pm0.0000001) \times E\\
 + 4.524(\pm3.13) \times 10^{-11} \times E^2.
\end{split}
\end{eqnarray}
\begin{eqnarray}
\begin{split}
\textrm{Min.I} = 2457467.73534(\pm0.00001) + 0.23287206(\pm0.00000001) \times E.
\end{split}
\end{eqnarray}
Equation (5) is for ASAS J124343+1531.7, which exists a long-term increase in the orbital period, and the rate is calculated as $dp/dt$ = 1.241 $\times$ 10$^{-7}$ $d\cdot yr^{-1}$. Equation (6) is for LINEAR 2323566, which obtained a new linear ephemeris. By comparison, it can be seen that the results of the two methods are almost the same.

\begin{figure}\centering
\includegraphics[width=0.44\textwidth]{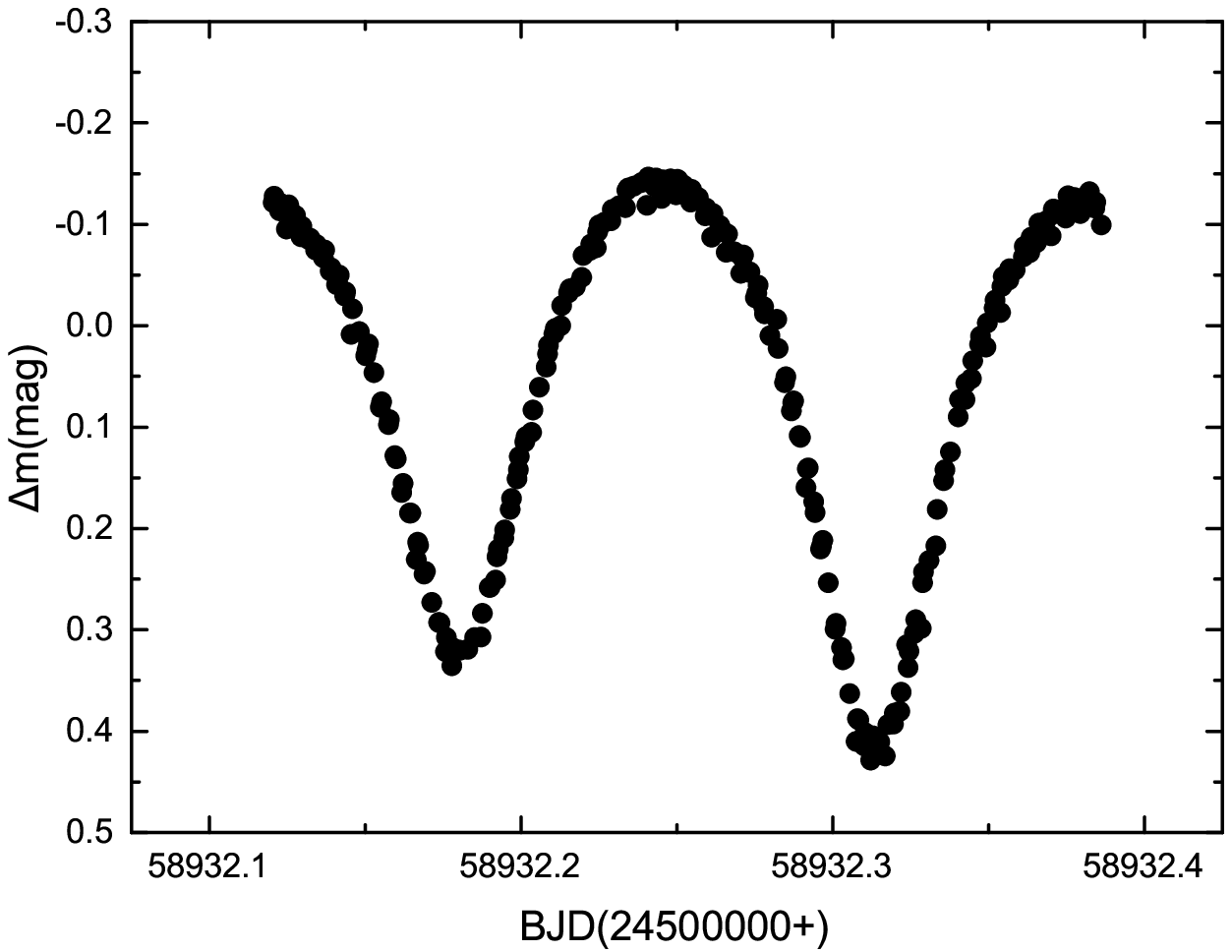}
\includegraphics[width=0.44\textwidth]{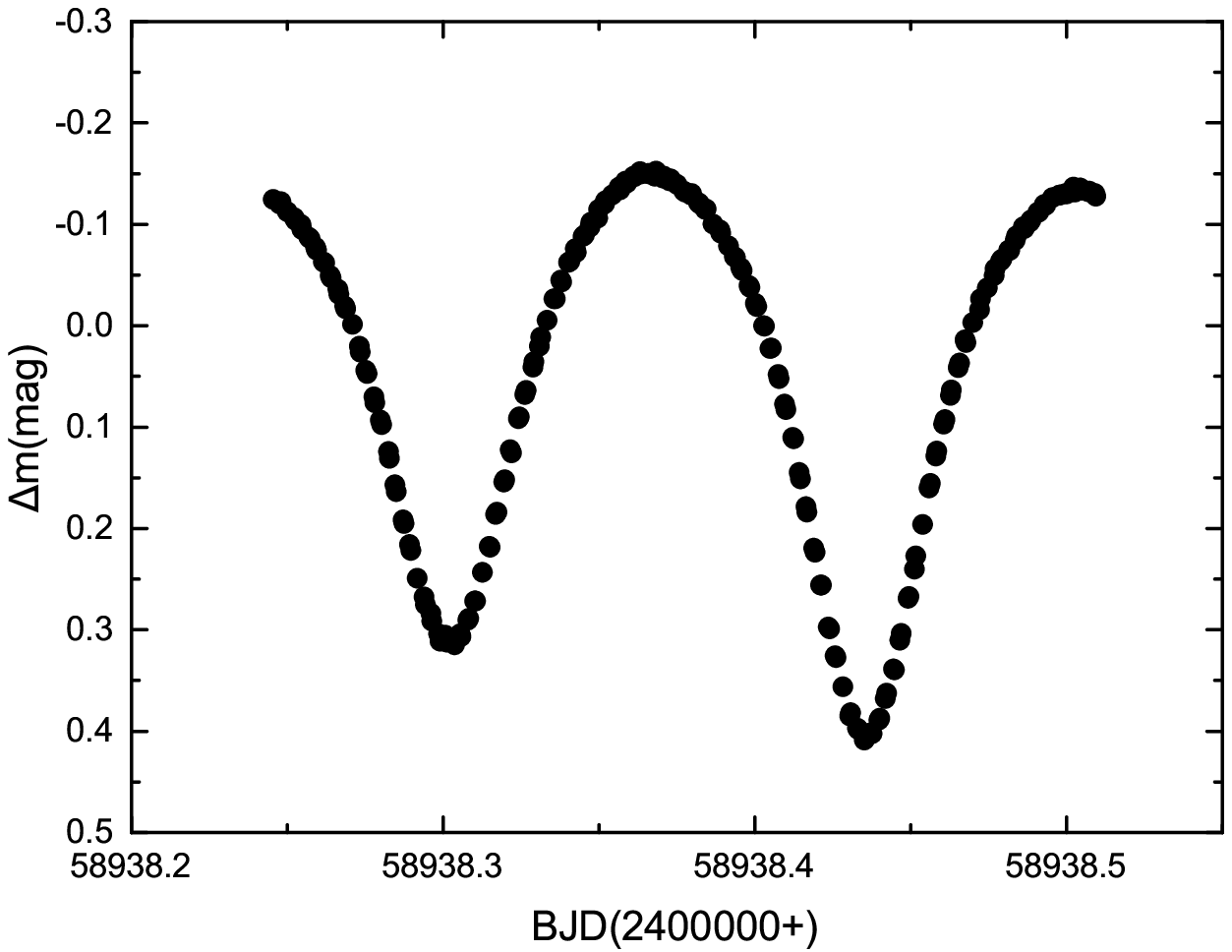}\\
\includegraphics[width=0.44\textwidth]{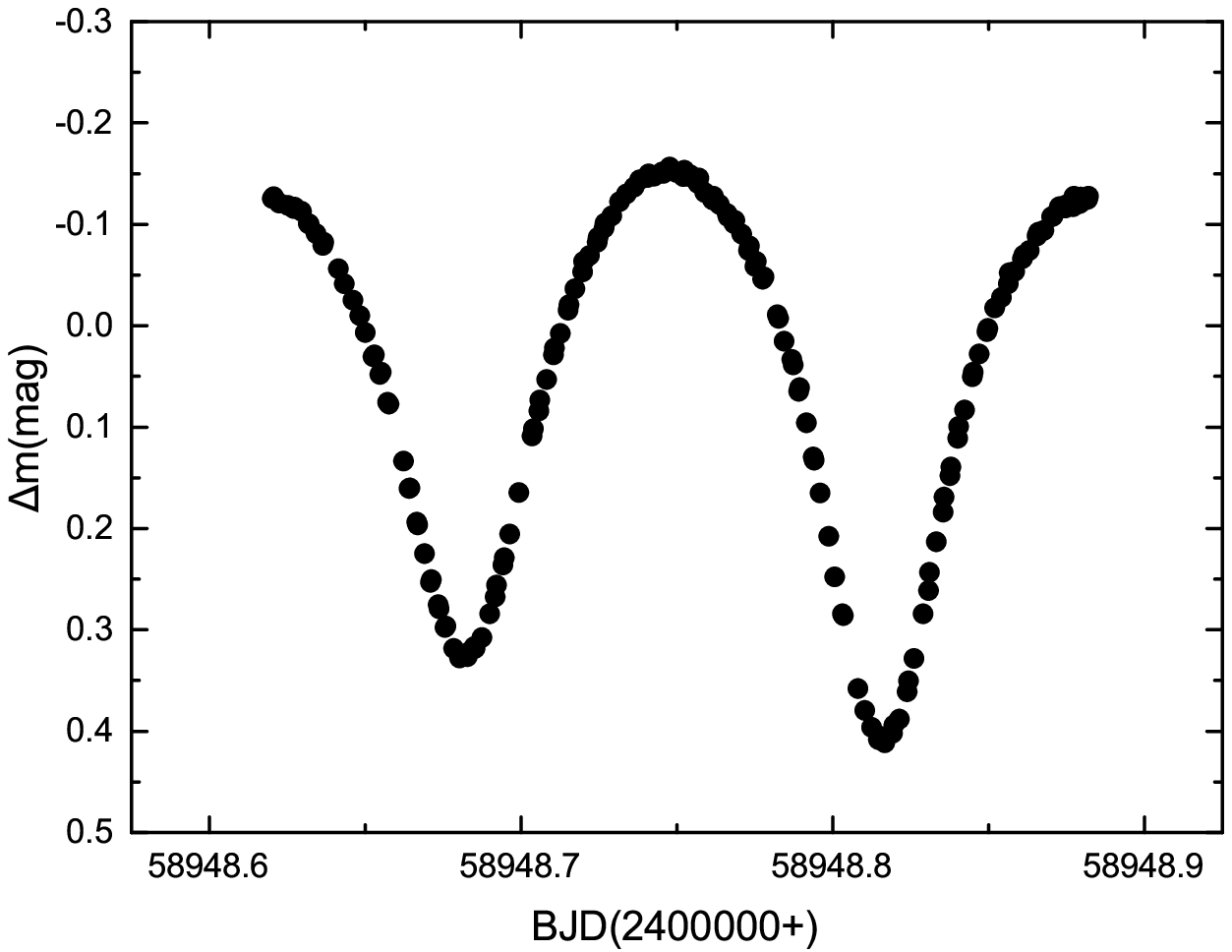}
\includegraphics[width=0.44\textwidth]{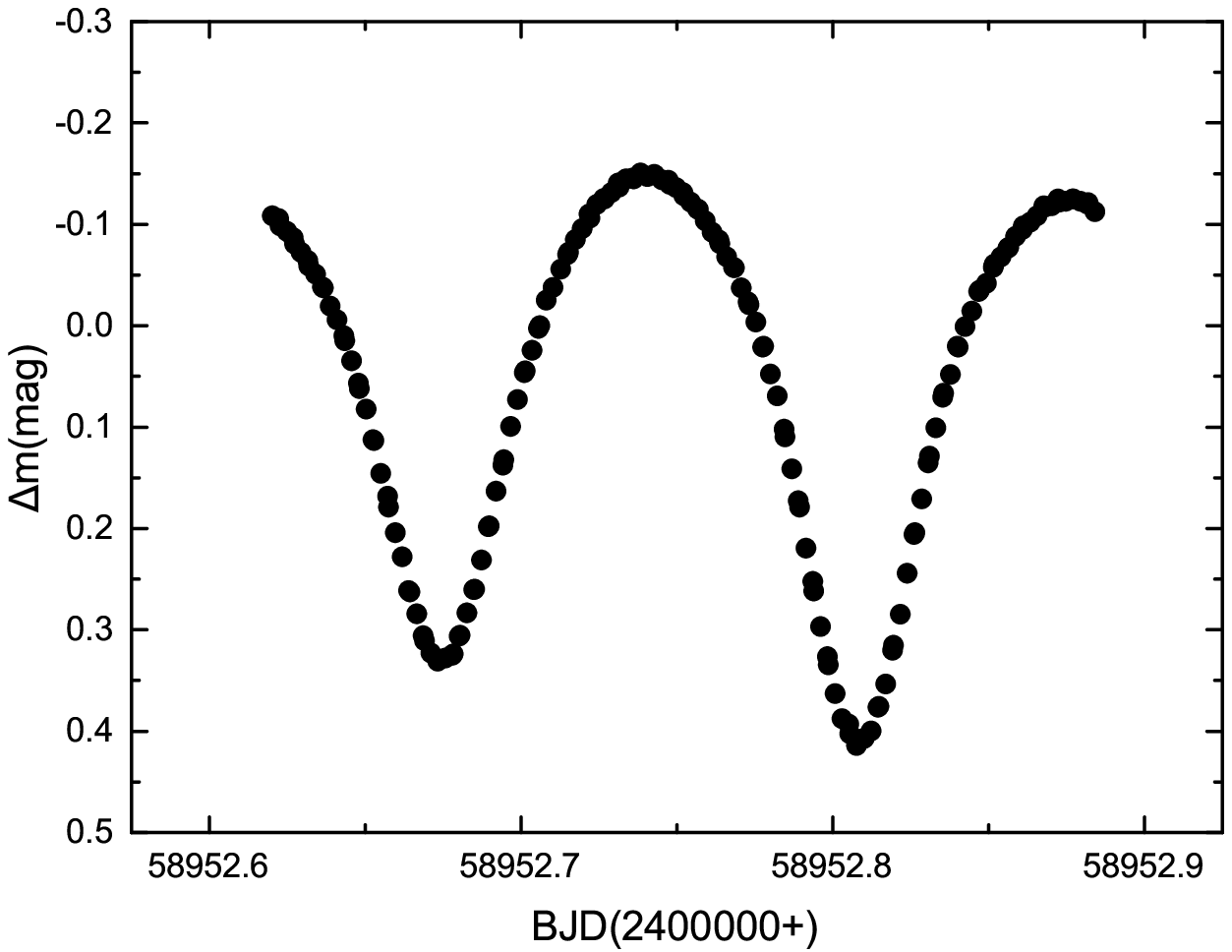}
\caption{The TESS light curves at different times of ASAS J124343+1531.7 after translation.}
\end{figure}

\begin{table*}[!htb]
\caption{Observed minimal timings and $O-C$ values of ASAS J124343+1531.7 and LINEAR 2323566}
\setlength{\leftskip}{-0pt}
\begin{center}
\begin{tabular}{lccccccc}
\hline\hline
 JD (Barycentric)2400000+ &  Error & Type  &   Min. &  Epoch  &  $O-C$     & Res. &  References \\
\hline
\multicolumn{8}{c}{ASAS J124343+1531.7}  \\
\hline
      54955.7337  & 0.0011 & ccd & s & -12304.5  &  0.0123    & -0.0002    & (1)  \\
      55280.7162  & 0.0003 & ccd & s & -11083.5  &  0.0110    & 0.0004     & (2)  \\
      55616.8773  & 0.0006 & ccd & s & -9820.5   &  0.0095    & 0.0007     & (3)  \\
      55983.9104  & 0.0006 & ccd & s & -8441.5   &  0.0052    & -0.0018    & (4)  \\
      56050.7191  & 0.0021 & ccd & s & -8190.5   &  0.0072    & 0.0006     & (4)  \\
      58230.7117  & 0.0002 & ccd & p & 0.0       &  0.0000    & 0.0003     & (5)  \\
      58508.0525  & 0.0002 & ccd & p & 1042.0    &  0.0000    & 0.0007     & (5)  \\
      58894.9169  & 0.0003 & ccd & s & 2495.5    &  -0.0021   & -0.0009    & (5)  \\
      58932.1801  & 0.0002 & ccd & s & 2635.5    &  -0.0015   & -0.0003    & (6)  \\
      58932.3137  & 0.0002 & ccd & p & 2636.0    &  -0.0011   & 0.0002     & (6)  \\
      58938.3020  & 0.0001 & ccd & s & 2658.5    &  -0.0014   & -0.0001    & (6)  \\
      58938.4355  & 0.0001 & ccd & p & 2659.0    &  -0.0010   & 0.0003     & (6)  \\
      58948.6821  & 0.0001 & ccd & s & 2697.5    &  -0.0017   & -0.0004    & (6)  \\
      58948.8160  & 0.0001 & ccd & p & 2698.0    &  -0.0008   & 0.0004     & (6)  \\
      58952.6745  & 0.0001 & ccd & s & 2712.5    &  -0.0017   & -0.0004    & (6)  \\
      58952.8084  & 0.0001 & ccd & p & 2713.0    &  -0.0008   & 0.0005     & (6)  \\
\hline
\multicolumn{8}{c}{LINEAR 2323566}  \\
\hline
      57467.7355  &0.0002 & ccd & p & 0.0    & 0.0000   & 0.0002   & (5)    \\
      57467.8515  &0.0001 & ccd & s & 0.5    & -0.0004  & -0.0002  & (5)    \\
      57467.9683  &0.0002 & ccd & p & 1.0    & -0.0001  & 0.0001   & (5)    \\
      58895.9397  &0.0003 & ccd & p & 6133.0 & -0.0084  & 0.0000   & (5)    \\
\hline
\end{tabular}
\end{center}
\scriptsize
Ref. (1)~\citet{Diethelm09}; (2)~\citet{Diethelm10}; (3)~\citet{Diethelm11}; (4)~\citet{Diethelm12};
(5) This paper; (6) TESS
\end{table*}

\begin{figure}\centering
\includegraphics[width=0.48\textwidth]{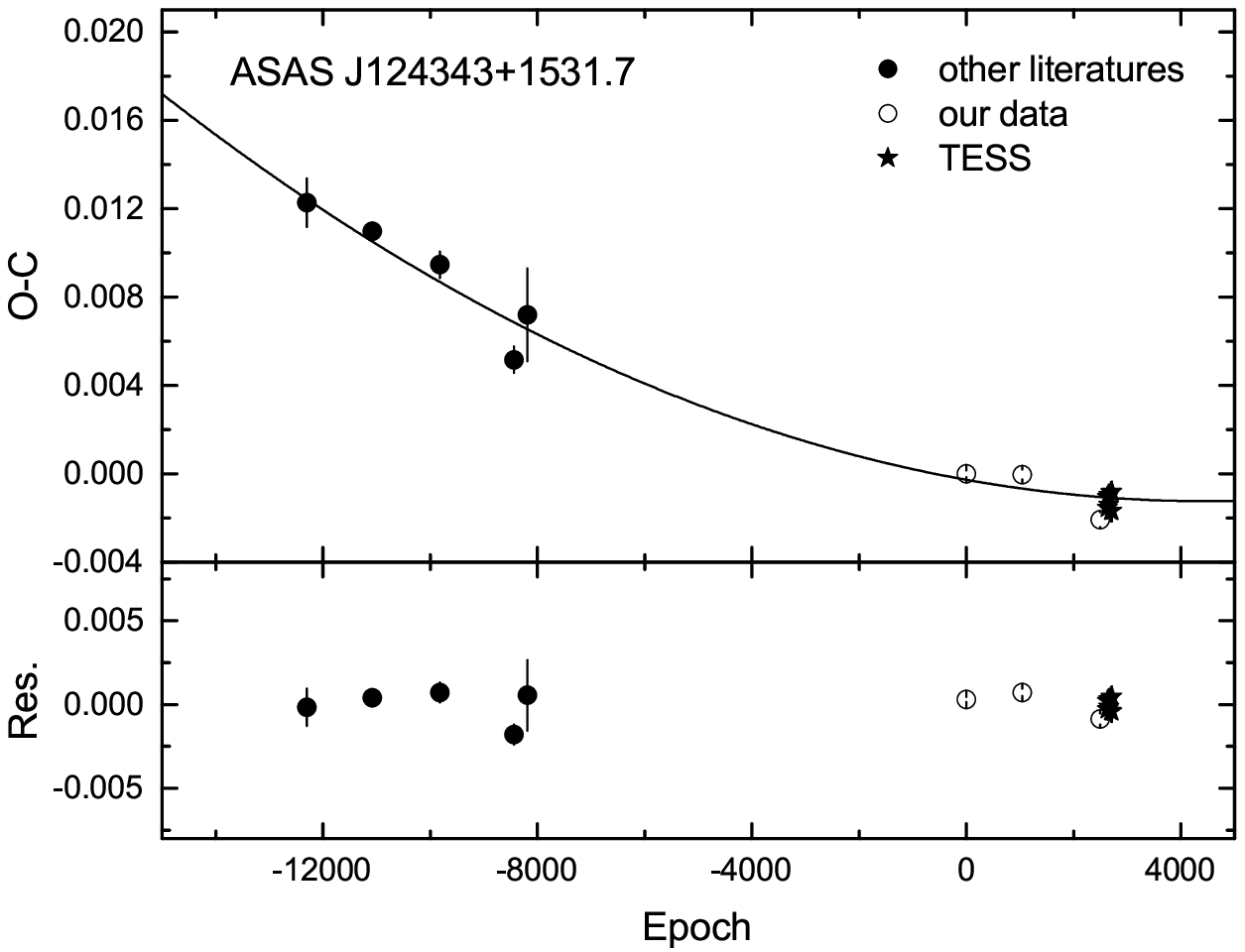}
\includegraphics[width=0.48\textwidth]{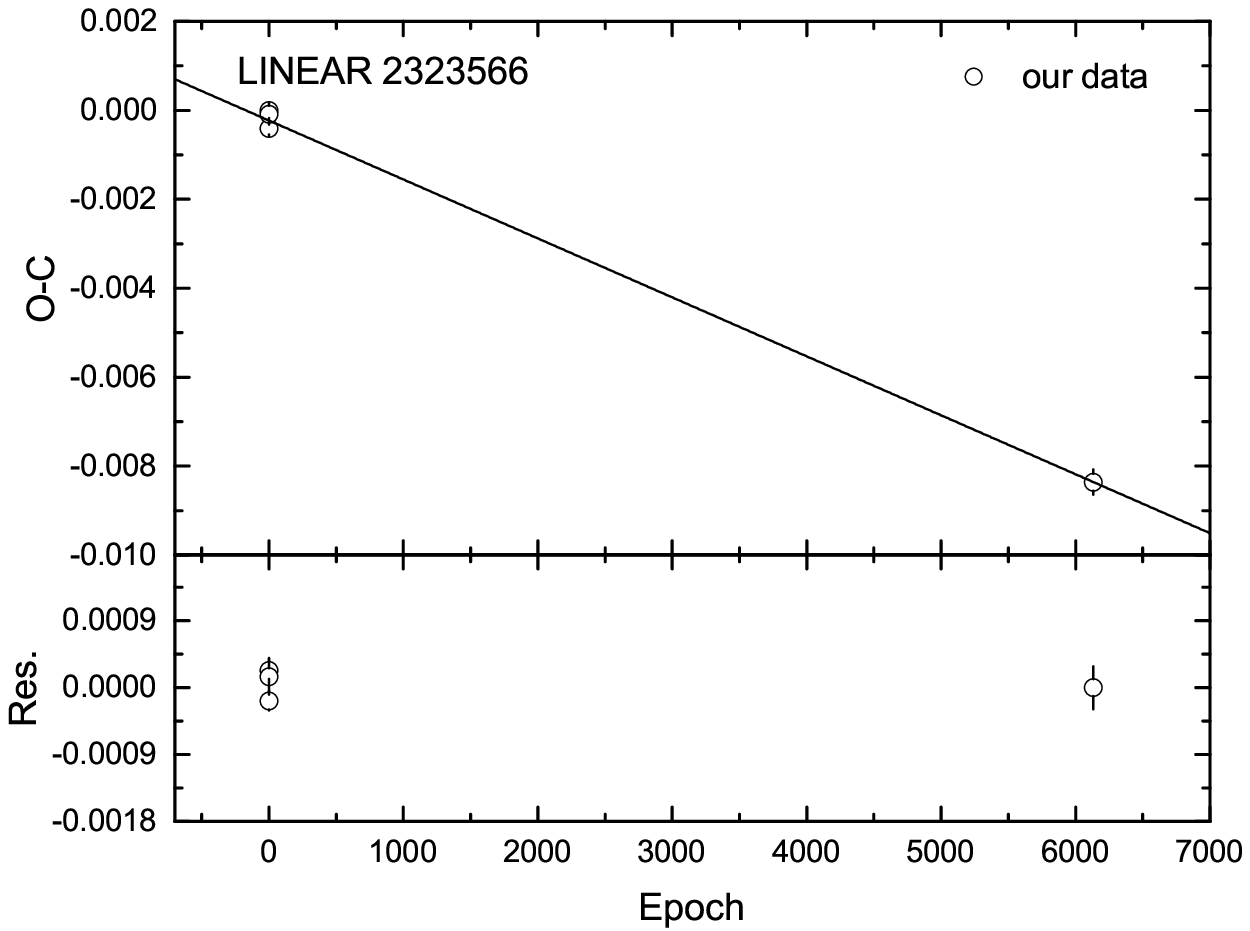}
\caption{$O-C$ diagram of ASAS J124343+1531.7 (left panel) and LINEAR 2323566 (right panel). Filled circles refer to the minima from other literatures, open black circles represent the minima from our observations and the filled pentagram represent the minima from TESS.}
\end{figure}

\section{Spectral analyses}
Since both ASAS J124343+1531.7 and LINEAR 2323566 exhibit the O'Connell effect, we used a spot generally caused by magnetic activity to explain this phenomenon. In order to further confirm whether in these two objects exist magnetic activity, we analyzed the spectral lines below. The observed spectra not only shows chromospheric activity, but also mix in some spectral lines of photosphere. In order to use the intensity of H$_\alpha$ line to analyze their magnetic activity, we adopted a spectral subtraction technique conducted by \citet{Barden85}. Some other objects also used this method, for example, ASAS J082243+1927.0 \citep{Kandulapati15}, EPIC 202073314 \citep{Sriram18}, V582 Lyr \citep{Cheng19}; V384 Ser, AQ Psc, V480 Gem and 2MASS J07095549+3643564 \citep{Zhang20}.

First, we selected some targets with a temperature difference of 100K above or below our objects from a catalogue of radial velocity standard stars \citep{Huang18}. Then we used these targets to search in LAMOST Data Release 7. Finally, we obtained 7 and 8 template spectra for ASAS J124343+1531.7 and LINEAR 2323566, respectively, and used these inactive spectra to construct the synthesized spectra. Combining the object and template spectra, a total of 18 spectra were normalized by IRAF package Continuum. We obtained their residuals by subtracting the comprehensive spectra from the object spectra, and all the spectra exclude the region of H$_\alpha$. Then the best template spectra were determined as 2MASS J19141861+4238492 and KIC 7446560 for the two objects respectively. The spectral type and temperature were determined as K1 and 5028K for 2MASS J19141861+4238492, K5 and 4377K for KIC 7446560. We subtracted their best template spectra from the observed spectra to obtain the subtracted spectra, which is the chromospheric construction. All of the observed, synthesized and subtracted spectra are displayed in Figure 9. From the intensity of emission line H$_\alpha$, LINEAR 2323566 has stronger chromospheric activity than ASAS J124343+1531.7. The equivalent widths (EWs) were calculated using IRAF package Splot, which are listed in Table 4.

\begin{figure}\centering
\includegraphics[width=0.90\textwidth]{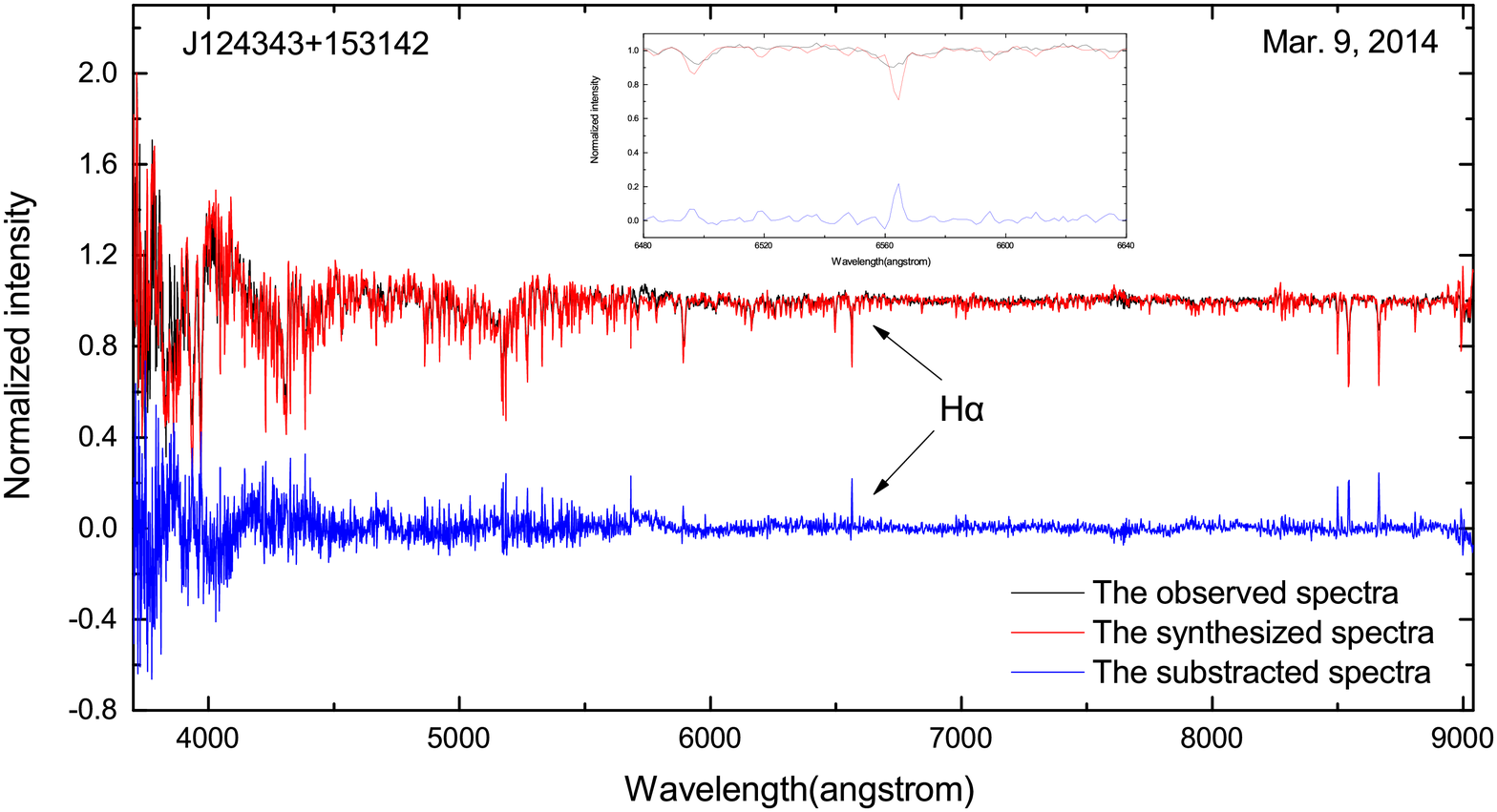}\\
\includegraphics[width=0.90\textwidth]{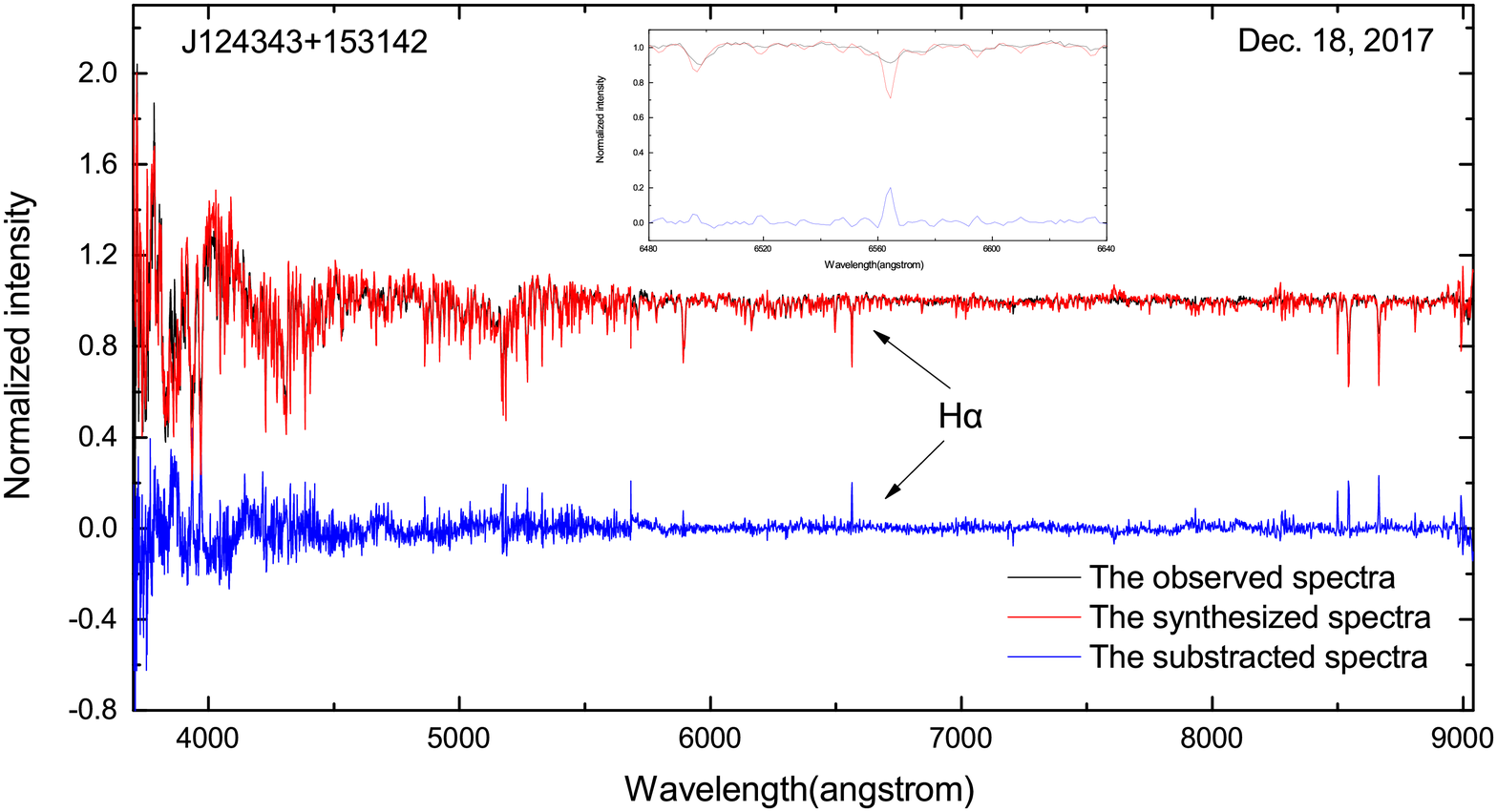}\\
\includegraphics[width=0.90\textwidth]{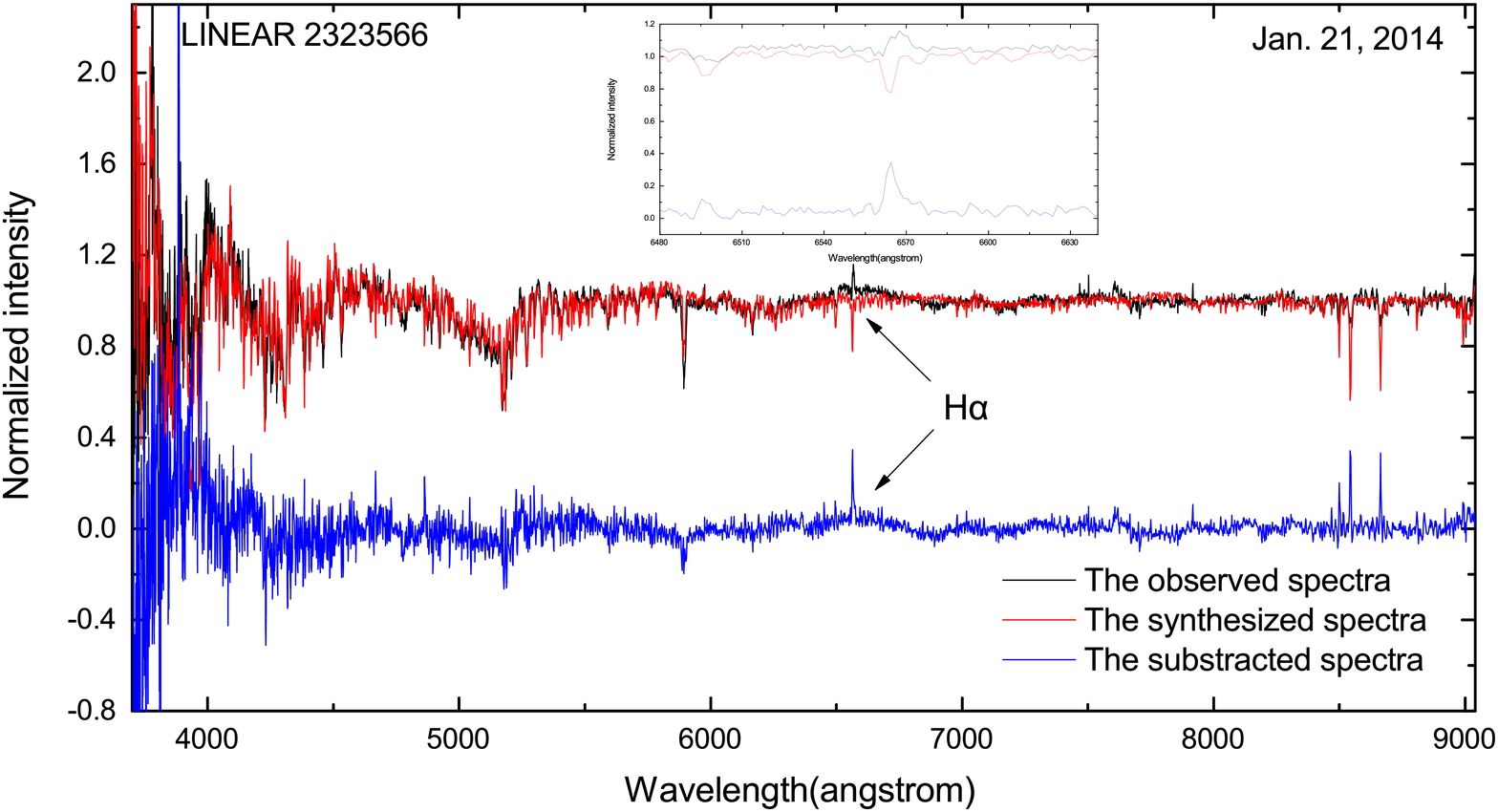}
\caption{Normalized observed, synthesized and subtracted spectra of ASAS J124343+1531.7 and LINEAR 2323566.}
\end{figure}

\section{Discussions and Conclusions}
Using the light curve analysis, we determined the physical parameters of ASAS J124343+1531.7 and LINEAR 2323566, which are shown in Table 5. The mass ratio and fill-out factor are 3.758 $\pm$ 0.028 and 31.8 $\pm$ 5.8 $\%$ for ASAS J124343+1531.7, 1.438 $\pm$ 0.040 and 14.9 $\pm$ 10.9 $\%$ for LINEAR 2323566. Therefore, ASAS J124343+1531.7 is a W-subtype median contact binary, while LINEAR 2323566 is a W-subtype shallow contact binary. The O'Connell effect of these two systems prove that it may exist magnetic activity. Spectral analysis obtained the EWs of the emission line  H$_\alpha$, which confirmed the possibility of magnetic activity of these two objects.

According to \citet{Yakut05}, the statistics show that the more massive stars of W-type and A-type low-temperature contact binaries (LTCBs) usually locate in the main sequence. Consequently, assuming the more massive star is a main sequence star and referring to \citet{Cox00}, the spectral type of the more massive star is classified by the effective temperature as K3 and K5, then the mass was determined by the spectral type as $M_2$ = 0.75$M_{\odot}$, 0.68$M_{\odot}$. Subsequently the mass of less massive star is also calculated according to the mass ratio and the mass of more massive star, $M_1$ = 0.20$M_{\odot}$, 0.47$M_{\odot}$. Combining the mass and period, the orbital semi-major axis was determined using the third law of Kepler, $M_1+M_2=0.0134\times a^3/P^2$, $a$ = 1.71$R_{\odot}$, 1.67$R_{\odot}$. Simultaneously, the radius and luminosity of the two components were computed by using the photometric solutions and Stefan-Boltzman law, $R_1$ = 0.49$R_{\odot}$, 0.60$R_{\odot}$, $R_2$ = 0.88$R_{\odot}$, 0.71$R_{\odot}$, $L_1$ = 0.15$L_{\odot}$, 0.13$L_{\odot}$, $L_2$ = 0.38$L_{\odot}$, 0.17$L_{\odot}$. The error bars are obtained by error transmission. All the absolute parameters of ASAS J124343+1531.7 and LINEAR 2323566 are listed in Table 8.

Both of the two targets have been observed by Gaia (\citealt{Gaia16,Gaia18}), we obtained the Gaia distances, 361.8 pc for ASAS J124343+1531.7 and 678.7 pc for LINEAR 2323566. Considering the interstellar extinction, we calculated their apparent magnitudes using luminosity from WD solution and Gaia distance, and obtained 13.536 mag for ASAS J124343+1531.7 and 15.838 mag for LINEAR 2323566. Our apparent magnitude of photometric values is about 13.20 mag for ASAS J124343+1531.7, while apparent magnitude from \citet{Drake14} is 15.38 mag for LINEAR 2323566. The apparent magnitudes derived by the two methods are somewhat different, this may be caused by that there are some differences between the estimated absolute parameters and the real ones.

\begin{table}
\begin{center}
\caption{Absolute parameters of ASAS J124343+1531.7 and LINEAR 2323566}
\begin{tabular}{lcccc}
\hline
            &  \multicolumn{2}{c}{ASAS J124343+1531.7}  &   \multicolumn{2}{c}{LINEAR 2323566}     \\
Components  &   Primary   &  Secondary  &  Primary   &  Secondary   \\
\hline
Mass ($M_{\odot}$)            &  0.20$\pm$0.01    &   0.75$\pm$0.02  &  0.47$\pm$0.03    &   0.68$\pm$0.02 \\
Radius ($R_{\odot}$)          &  0.49$\pm$0.01    &   0.88$\pm$0.01  &  0.60$\pm$0.01    &   0.71$\pm$0.03   \\
Luminosity ($L_{\odot}$)      &  0.15$\pm$0.02    &   0.38$\pm$0.06  &  0.13$\pm$0.02    &   0.17$\pm$0.03  \\
Semi-major axis ($R_{\odot}$) &  \multicolumn{2}{c}{1.71$\pm$0.01}    &   \multicolumn{2}{c}{1.67$\pm$0.03}  \\
\hline
\end{tabular}
\end{center}
\end{table}

\subsection{Mass transfer between two components}
According to the analysis of the orbital period, the overall trend of $O-C$ shows an upward parabolic composition with a rate of $dp/dt$ = 1.239 $\times$ 10$^{-7}$ $d yr^{-1}$ for ASAS J124343+1531.7. The long-term increase can probably be explained by the mass transfer from the less massive to the more massive star. Then we used the following equation to calculate the rate of mass transfer,
\begin{eqnarray}
{\dot{P}\over P}=-3\dot{M_1}({1\over M_1}-{1\over M_2}),
\end{eqnarray}
$dM_1/dt=-4.22\times10^{-8}\,M_\odot$ yr$^{-1}$ was determined. The negative sign indicates that the less massive star $M_1$ is losing mass, while the more massive star $M_2$ is receiving mass. Then we discuss the possible subsequent evolution of ASAS J124343+1531.7. As the mass transfer, mass ratio will increase and the distance between the two components will increase. Therefore, the degree of contact will gradually become smaller, ASAS J124343+1531.7 will evolve from the current contact binary to a semi-contact binary or detached binary. Consequently, ASAS J124343+1531.7 is a good target for studying the thermal relaxation oscillation (TRO) model \citep{Lucy76,Flannery76,Robertson77}, follow-up observations are also needed to confirm this result.

\subsection{Evolutionary status}
In order to study the evolutionary status of these two objects, the diagram of mass-luminosity and mass-radius were plotted in Figure 10. The solid and dotted lines show the zero age main sequence (ZAMS) and the terminal age main sequence (TAMS) constructed by the BSE Code ~\citep{Hurley02}. Circles and rhombuses respectively represent 30 A-subtype and 42 W-subtype binary derived from~\citet{Yakut05}. Red stars and blue stars respectively represent the two components of ASAS J124343+1531.7 and LINEAR 2323566. All solid symbols represent the more massive components, whereas the open symbols denote the less massive ones. The positions of the two objects in the diagram are consistent with the trend of the other binaries. The more massive stars of ASAS J124343+1531.7 and LINEAR 2323566 locate above the ZAMS line and below the TAMS line. The less massive star of ASAS J124343+1531.7 locates above the TAMS line, while LINEAR 323566 locates above the ZAMS line and below the TAMS line. This manifests that the two components of LINEAR 2323566 are both slightly evolved stars. And for ASAS J124343+1531.7, the more massive star is a main sequence star, the less massive star has evolved out of the main sequence and is over-luminous and over-sized.

In summary, the photometric solutions and period variation of ASAS J124343+1531.7 and LINEAR 2323566 were obtained first. Through light curve analysis, it is found that ASAS J124343+1531.7 is a W-subtype median contact binary, while LINEAR 2323566 is a W-subtype shallow contact binary. They both present the O'Connell effect, which can be explained by a cool spot on the primary star. The existence of the cool spot may be caused by magnetic activity, and at the same time, the weak H$\alpha$ emission lines on the spectra also indicate that the two objects possess magnetic activity. Through the analysis of the orbital period, we obtained ASAS J124343+1531.7 exists a secular period increase, which is likely caused by the mass transfer from the less massive star to the more massive star. We obtained a new ephemeris for LINEAR 2323566. The two components of ASAS J124343+1531.7 are both main sequence stars. While for LINEAR 2323566, the more massive star is a main sequence star, the less massive star has evolved out of main sequence and is over-luminous and over-sized. In the future, more observations combining photometry and spectra are needed to verify the variations of orbital period and the evolution of ASAS J124343+1531.7 and LINEAR 2323566.

\begin{figure}\centering
\includegraphics[width=0.45\textwidth]{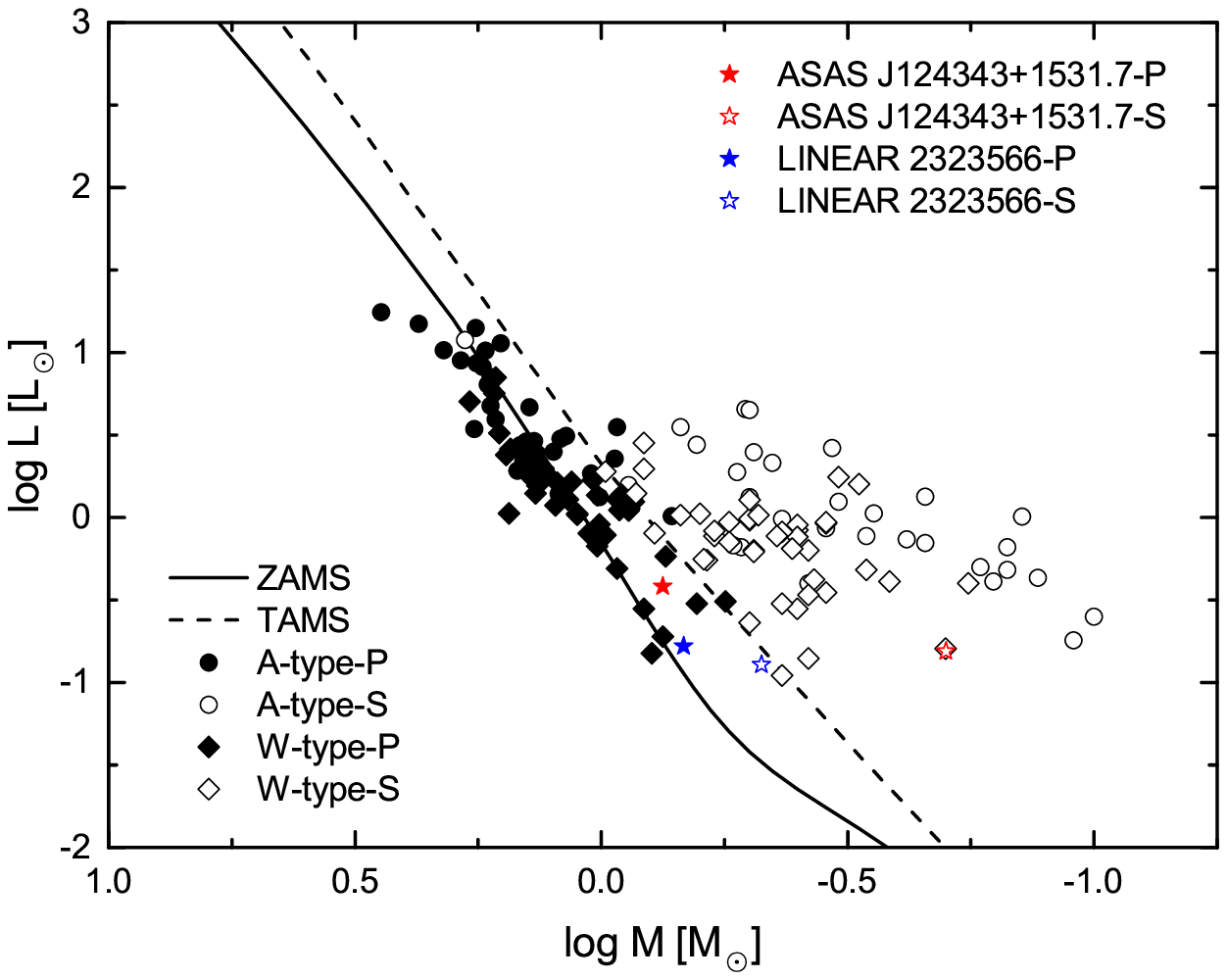}
\includegraphics[width=0.46\textwidth]{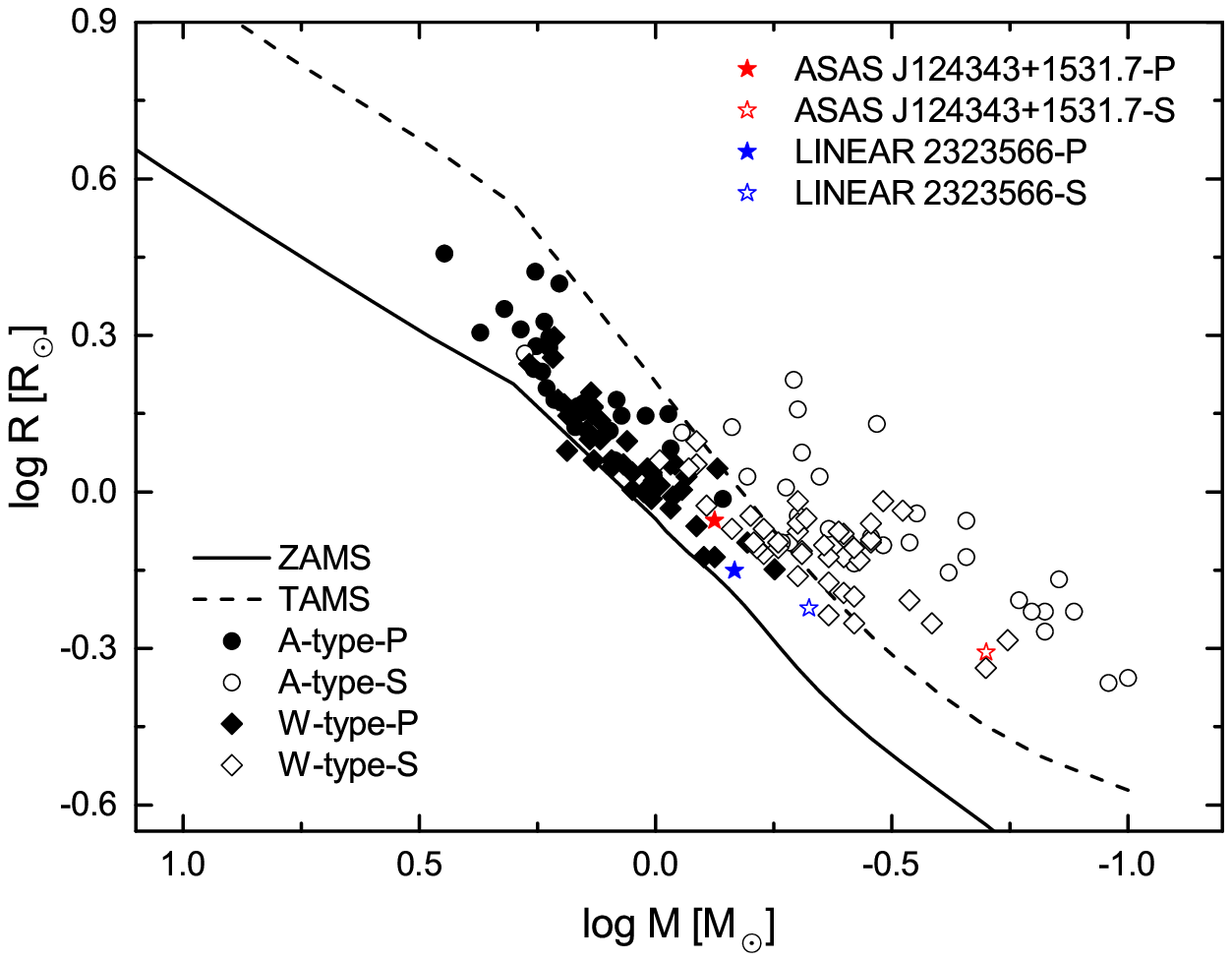}
\caption{Mass-luminosity diagram (left panel) and mass-radius diagram (right panel). Circles and rhombuses respectively represent 30 A-subtype and 42 W-subtype binary derived from~\citet{Yakut05}. Red stars and blue stars respectively represent the two components of ASAS J124343+1531.7 and LINEAR 2323566. All solid symbols represent the more massive components, whereas the open symbols denote the less massive ones.}
\end{figure}

\acknowledgments{We are very grateful for the very helpful comments and suggestions of the referee. This work is supported by the Joint Research Fund in Astronomy (No. U1931103) under cooperative agreement between NSFC and Chinese Academy of Sciences (CAS), and by National Natural Science Foundation of China (NSFC) (No. 11703016), and by Young Scholars Program of Shandong University, Weihai (Nos. 20820171006), and by the Open Research Program of Key Laboratory for the Structure and Evolution of Celestial Objects (No. OP201704). RM acknowledges the financial support from the Universidad Nacional Aut\'{o}noma de M\'{e}xico (UNAM) and DGAPA under grant PAPIIT IN 100918.

We acknowledge the support of the staff of OAN-SPM. Guoshoujing Telescope (the Large Sky Area Multi-Object Fiber Spectroscopic Telescope LAMOST) is a National Major Scientific Project built by the Chinese Academy of Sciences. Funding for the project has been provided by the National Development and Reform Commission. LAMOST is operated and managed by the National Astronomical Observatories, Chinese Academy of Sciences. This paper includes data collected by the $TESS$ mission, which are publicly available from the Mikulski Archive for Space Telescope (MAST). Funding for the $TESS$ mission is provided by NASA's Science Mission directorate. We acknowledge the $TESS$ team for its support of this paper. This work has made use of data from the European Space Agency (ESA) mission Gaia (https://www.cosmos.esa.int/gaia), processed by the Gaia Data Processing and Analysis Consortium (DPAC; https: //www.cosmos.esa.int/web/gaia/dpac/consortium). Funding for the DPAC has been provided by national institutions, in particular the institutions participating in the Gaia Multilateral Agreement.

The calculations in this work were carried out at Supercomputing Center of Shandong University, Weihai.}

\end{document}